\documentclass[12pt]{iopart}
\usepackage{iopams} 
\usepackage{isolatin1}
\usepackage{graphicx}
\usepackage[german]{varioref}
\usepackage[dvips]{epsfig}
\usepackage{graphics}
\usepackage[dvips]{color}
\usepackage{color}
\usepackage{colordvi}
\usepackage{dsfont}
\usepackage{psfrag}
\begin{document}
\newcommand{\be}{\begin{eqnarray}}
\newcommand{\ee}{\end{eqnarray}}
\def\p#1#2{|#1\rangle \langle #2|}
\def\ket#1{|#1\rangle}
\def\bra#1{\langle #1|}
\def\refeq#1{(\ref{#1})}
\def\tb#1{{\overline{{\underline{ #1}}}}}
\def\im{\mbox{Im}}
\def\re{\mbox{Re}}
\def\nn{\nonumber}
\def\t{\mbox{tr}}
\def\sgn{\mbox{sgn}}
\def\Li{\mbox{Li}}
\def\P{\mbox{P}}
\def\d{\mbox d}
\def\i{\int_{-\infty}^{\infty}}
\def\ip{\int_{0}^{\infty}}
\def\mi{\int_{-\infty}^{0}}
\def\A{\mathfrak A}
\def\AA{{\overline{{\mathfrak{A}}}}}
\def\a{\mathfrak a}
\def\aa{{\overline{{\mathfrak{a}}}}}
\def\B{\mathfrak B}
\def\BB{{\overline{{\mathfrak{B}}}}}
\def\b{\mathfrak b}
\def\bb{{\overline{{\mathfrak{b}}}}}
\def\R{\mathcal R}
\def\dm{\mathfrak d}
\def\dd{{\overline{{\mathfrak{d}}}}}
\def\D{\mathfrak D}
\def\DD{{\overline{{\mathfrak{D}}}}}
\def\c{\mathfrak c}
\def\cc{{\overline{{\mathfrak{c}}}}}
\def\C{\mathfrak C}
\def\CC{{\overline{{\mathfrak{C}}}}}
\def\O{\mathcal O}
\def\F{\mathcal F_k}
\def\N{\mathcal N}
\def\I{\mathcal I}
\def\S{\mathcal S}
\def\P{\mathcal P}

\def\G{\Gamma}
\def\L{\Lambda}
\def\la{\lambda}
\def\g{\gamma}
\def\al{\alpha}
\def\s{\sigma}
\def\e{\epsilon}
\def\te{{\rm e}}
\def\max{{\rm max}}
\def\str{{\rm str}}
\def\tC{\text C}
\def\Fo{\mathcal{F}_{1,k}}
\def\Ft{\mathcal{F}_{2,k}}
\def\vs{\varsigma}
\def\l{\left}
\def\r{\right}
\def\up{\uparrow}
\def\down{\downarrow}
\def\u{\underline}
\def\ov{\overline}
\title[Lattice path integral approach to the Kondo model]{Lattice path integral approach to the one-dimensional
  Kondo model}

\author{Michael Bortz\footnote{Email: bortz@physik.uni-wuppertal.de} and Andreas Kl\"umper
}

\address{Bergische Universit\"at Wuppertal, Fachbereich Physik, Theoretische
  Physik, \\ 42097 Wuppertal, Germany}
\begin{abstract}
An integrable Anderson-like impurity model in a correlated host is derived from a gl(2$|$1)-symmetric transfer matrix by means of the Quantum-Inverse-Scattering-Method (QISM). Using the Quantum Transfer Matrix
technique, free energy contributions of both the bulk and the impurity are
calculated exactly. As a special case, the limit of a localized moment in a free bulk (Kondo limit)
is performed in the Hamiltonian and in the free energy. In this case, high- and
low-temperature scales are calculated with high accuracy. 
\end{abstract}


\submitto{\JPA}


\section{Introduction}
Over decades, the model of a localized magnetic impurity in a non-magnetic
metal has been one of the major challenges in many-particle theory. Anderson \cite{and61}
proposed a model of a localized impurity interacting with a host of free
electrons through hybridization:
\be
H_A&=& \sum_{k,\tau} \e_kc^\dag_{k,\tau} c_{k,\tau} +\e_d n_d+\sum_{k,\tau}
\l(M_kd^\dag_\tau c_{k,\tau}+M^*_kc^\dag_{k,\tau}d_\tau\r)+Un_{d,\up} n_{d,\down}\nn
\ee
On the impurity site, a Coulomb repulsion $U$ is allowed. The scattering at the impurity is assumed to be
isotropic and therefore one-dimensional. In the limit $|M_k|^2/U\ll 1$, a
localized moment forms, which is demonstrated by a canonical transformation of the
Anderson model, resulting in the Kondo model \cite{sch66} with an impurity
operator
\be
H_i&=& 2J\sum_{k,k'}c^\dag_{k,\tau} \bsigma_{\tau,\tau'} c_{k',\tau'}
\bsigma_i\label{konmod}\;,
\ee
where $\bsigma=(\sigma^x,\sigma^y,\sigma^z)$ denotes the Pauli matrices.
The Kondo model
describes a free host, interacting weakly with a localized magnetic moment via
antiferromagnetic XXX spin
exchange with an amplitude $J$. ``Weak interaction'' means that at high
temperatures, the coupling is negligible and the impurity spin shows
Curie-Weiss behavior. The lesson to be
learned from this limit is that a localized moment occurs if singly occupied sites are energetically favourable and hybridization only 
leads
to virtual double or zero occupation. The model \refeq{konmod} served
as starting point for Kondo \cite{kon64}, who performed a perturbational
calculation of the scattering amplitude between host and impurity up to third
order in $J$. He discovered a $\ln \widetilde T_K/T$ contribution to the
electrical resistivity. $\widetilde T_K$ is the crossover
temperature which indicates the limit of perturbation theory: A divergence
occurs for $T\sim \widetilde T_K$. 

The method which overcomes the failure of perturbation theory is
scaling. By the
implementation of his numerical renormalization group, Wilson \cite{wil75}
achieved a non-perturbative calculation of the impurity contribution to the magnetic susceptibility $\chi$ and the
specific heat $C$ at $T\ll \widetilde T_k$. By assuming a linear dispersion in
the conduction band he discovered Fermi-liquid-like behavior of the impurity for $T\ll \widetilde T_K$. In the other extreme, at $T\gg T_K$, Wilson confirmed the asymptotic expansion in $\ln \widetilde T_K/T$ discovered by perturbation theory techniques.

Andrei
\cite{andr80} and Wiegmann \cite{wie81} obtained the spectrum
exactly by the Bethe Ansatz (BA). The linearized energy-momentum relation turned
out to be crucial for the application of the BA. Thermodynamic equilibrium
response functions were calculated in the following by employing thermodynamic
BA (TBA) techniques, \cite{tsv83}. The impurity contribution to the free energy is encoded in a set of
infinitely many coupled NLIE. These contain
the whole information about thermodynamic equilibrium functions. Especially, the
asymptotic high-temperature expansion due to Kondo and the Wilson ratios are encoded therein. The low temperature Fermi-liquid-like behavior was confirmed in the cited works. However, the high-temperature asymptotic expansion was not performed as far as in Wilson's approach. 

We develop a lattice path integral representation of the free energy of a
one-dimensional Anderson-like impurity model in a correlated host. This model can be viewed as a
lattice-regularized version of the Anderson model in the continuum. As a special case, the Kondo model ist obtained in a certain scaling limit. The host is based on a
four-dimensional representation of the Lie superalgebra gl(2$|$1). The corresponding four states
per lattice site are zero, single (with spin up or down) and double
occupation. The impurity degrees of freedom are
described by a three-dimensional representation of gl(2$|$1). Double occupation on the impurity site is excluded from the
beginning. The parameters of the model can be tuned such that on one hand,
particle exchange between the impurity site and adjacent host sites is eliminated by a canonical transformation and on the other hand zero occupation is energetically
suppressed. The same parameter tuning makes vanish correlations in the host. These conditions fulfilled, a localized moment in a free host occurs (Kondo limit). 

In order to regularize the continuous
Kondo model, it is quite natural to choose the
superalgebra gl(2$|$1).  Its even subalgebra is u(1)$\otimes$su(2), encoding
charge and spin degrees of freedom, respectively. Spin-charge separation
occurs in one dimension for interacting electron systems \cite{hal81}, and the impurity is
supposed to possess exclusively spin degrees of freedom. Indeed, the scaling limit reduces gl(2$|$1) to one of its subalgebras, su(2), in the impurity space. Then the excitation spectrum contains only spin degrees of freedom on the impurity site. 

The model proposed in this work allows for the Kondo limit as one special case. We calculate the free energy exactly in the general case, that is an Anderson-like impurity in an interacting host and perform the Kondo limit afterwards. Thus our results are farther reaching than the known non-perturbative treatments of the Kondo model \cite{wil75, tsv83}. The free energy of the host and of
the impurity are given by eigenvalues of distinct quantum
transfer matrices and can therefore be separated. In the Kondo limit, Wilson's results are confirmed with high accuracy. The general Anderson-like case will be investigated elsewhere \cite{akb}.

This article is organized as follows. In the next section, we
derive the Hamiltonian by QISM. The third section deals with the calculation of the
free energy. In each of these sections, the Kondo limit is treated
explicitly. Section four contains the derivation of Wilson's results in the framework of our path integral approach. A conclusion and an outlook form the last section. 

In all what follows we set $k_B=1$, and $g\mu_B=1$, where $k_B$ is Boltzmann's
constant, $g$ is the gyromagnetic factor and $\mu_B$ is the Bohr magneton. An
index $i$ ($h$) denotes quantities pertaining to the impurity (host). 

\section{The impurity model}
Let $V^{(d)}$ be the module giving rise to the $d$-dimensional irrep of
gl(2$|$1), $d=3,4$. A grading is assigned to the basis vectors through the
parity function $p$,
\be
\begin{array}{lll}
d=4\,:&p[1]=p[4]=0\,;& p[2]=p[3]=1 \\
d=3\,:&p[1]=p[2]=0\,;& p[3]=1\; .
\end{array}\label{pardef}
\ee
The matrices $R_{i,j}^{(d,d')}(u)\in$End$\l(V_i^{(d)}\otimes V_j^{(d')}\r)$
satisfy the graded Yang-Baxter-Equation (YBE),
\be
\fl \l[R_{2,3}^{(d,d')}(u)\r]^{\beta, \g}_{\beta', \g'}\, \l[R_{1,2}^{(d'',d')}(v)\r]^{\alpha, \g'}_{\alpha', \g''}\,
\l[R_{1,3}^{(d'',d)}(v-u)\r]^{\alpha', \beta'}_{\alpha'',
  \beta''}\,(-1)^{(p[\alpha]+p[\alpha'])p[\beta']}\nn\\
\fl\qquad =\l[R_{1,3}^{(d'',d)}(v-u)\r]^{\alpha, \beta}_{\alpha',\beta'}\, \l[R_{1,2}^{(d'',d')}(v)\r]^{\alpha', \g}_{\alpha'', \g'}\,
\l[R_{2,3}^{(d,d')}(u)\r]^{\beta', \g'}_{\beta'',
  \g''}(-1)^{(p[\alpha']+p[\alpha''])p[\beta']}\label{ybedef}\; .
\ee
Summation over doubly occurring indices is implied in the foregoing equation
and in all what follows. 

Explicit expressions of the $R$ matrices are given in the following, 
\be
R^{(3,3)}(u)&=& \frac{1}{u+1}\l(u+(-1)^{p[a]p[b]}e_a^b\otimes e_b^a\r)\label{defr33}\\
R^{(3,4)}(u)&=&
\frac{1}{u+\frac{\alpha}{2}+1}\l(u+\frac{\alpha}{2}+1+(-1)^{p[a]p[b]}e_a^b\otimes
E_b^a\r)\label{defr34}\\
R^{(4,4)}(u)&=& -\l(1+\frac{2\alpha}{u-\alpha}\check
P_1 - \frac{2\alpha +2}{u+\alpha +1}\check P_3\r)\label{defr44}\; .
\ee
$e_a^b$ ($E_a^b$) are the nine three- (four-) dimensional generators of
gl(2$|$1), obeying 
\be
\l[e^a_b,e^c_d\r]_{\pm}&:=& e^a_b\,e^c_d
-(-1)^{(p[a]+p[b])(p[c]+p[d])}e^c_d\,e^a_b\nn\\
&=& \delta^a_d\,e^c_b-(-1)^{(p[a]+p[b])(p[c]+p[d])} \delta^c_b\,e^a_d\label{defcom}\;,
\ee
and the same for the $E^a_b$. $\check P_1, \,\check P_3$ are projectors from
$V^{(4)}\otimes V^{(4)}$ onto gl(2$|$1) modules with highest weights
$(0,0|2\al)$ and $(-1,-1|2\al+2)$ respectively. They are given explicitly in \cite{bra95}.
For a matrix representation of $e^a_b$, choose the basis
\be
|\ov 1\rangle=(1,0,0)\;, \;|\ov 2\rangle=(0,1,0)\;,\;|\ov 3\rangle=(0,0,1)\nn\;.
\ee
Then $e^a_b:=|\ov b\rangle\langle \ov a|$ is the usual matrix
representation of projectors in three dimensional space. 

As to the $E^a_b$, we choose a basis in $V^{(4)}$, 
\be
|1\rangle=(1,0,0,0)\;,\;
| 2\rangle=(0,1,0,0)\;,\; |3\rangle=(0,0,1,0)\;,\;|4\rangle=(0,0,0,1)\nn\; .
\ee
We call the projectors associated with these states $m_b^a:=|b\rangle\langle a|$, $a,b=1,2,3,4$. One verifies
that the set \cite{bra95}
\begin{equation}
\eqalign{
E^1_1&=-\p{3}{3}-\p{4}{4},\qquad E^2_2=-\p{2}{2}-\p{4}{4},\nn\\
E^3_3&= \alpha\p{1}{1} +(\alpha+1)(\p{2}{2}+\p{3}{3})+(\alpha+2)\p{4}{4},\nn\\
E^2_1&=\p{2}{3},\qquad E^1_2=\p{3}{2},\nn\\
E^3_2&=\sqrt{\alpha}\p{1}{2}+\sqrt{\alpha+1}\p{3}{4},\qquad
E^2_3=\sqrt{\alpha}\p{2}{1}+\sqrt{\alpha+1}\p{4}{3},\nn\\
E^3_1&= -\sqrt{\alpha} \p{1}{3}+\sqrt{\alpha+1}\p{2}{4},\qquad E^1_3=
-\sqrt{\alpha} \p{3}{1}+\sqrt{\alpha+1}\p{4}{2}}\label{fourdimset}
\end{equation}
satisfies eq. \refeq{defcom}. In the sequel, the real parameter $\al$
is restricted to $\al>0$. 

Consider the set of matrices $\ov R$ defined by
\be
\l[\ov
R^{(d',d)}(u)\r]^{\alpha,\beta}_{\gamma,\delta}=(-1)^{p[\delta](p[\g]+p[\alpha])}
\l[ R^{(d',d)}(-u)\r]^{\g,\beta}_{\al,\delta}\label{rbardef}\; .
\ee
The permutation of the indices means that creators and
annihilators are exchanged in the auxiliary space of $R$. These $\ov R$-matrices satisfy 
\be
\fl\l[\ov R^{(d,d')}(u)\r]^{\beta, \g}_{\beta', \g'}\, \l[\ov R^{(d'',d')}(v)\r]^{\alpha, \g'}_{\alpha', \g''}\,
\l[R^{(d'',d)}(v-u)\r]^{\alpha', \beta'}_{\alpha'',
  \beta''}\,(-1)^{(p[\alpha]+p[\alpha'])p[\beta']}\nn\\
\fl\qquad=\l[R^{(d'',d)}(v-u)\r]^{\alpha, \beta}_{\alpha',\beta'}\, \l[\ov R^{(d'',d')}(v)\r]^{\alpha', \g'}_{\alpha'', \g''}\,
\l[\ov R^{(d,d')}(u)\r]^{\beta', \g'}_{\beta'',
  \g''}(-1)^{(p[\alpha']+p[\alpha''])p[\beta']}\label{ybebardef}\; .
\ee

The $R$-matrices can be translated into a graphical language. Straight lines
denote the 4-dimensional space, wavy lines symbolize three-dimensional space,

\begin{center}
\includegraphics[scale=0.5]{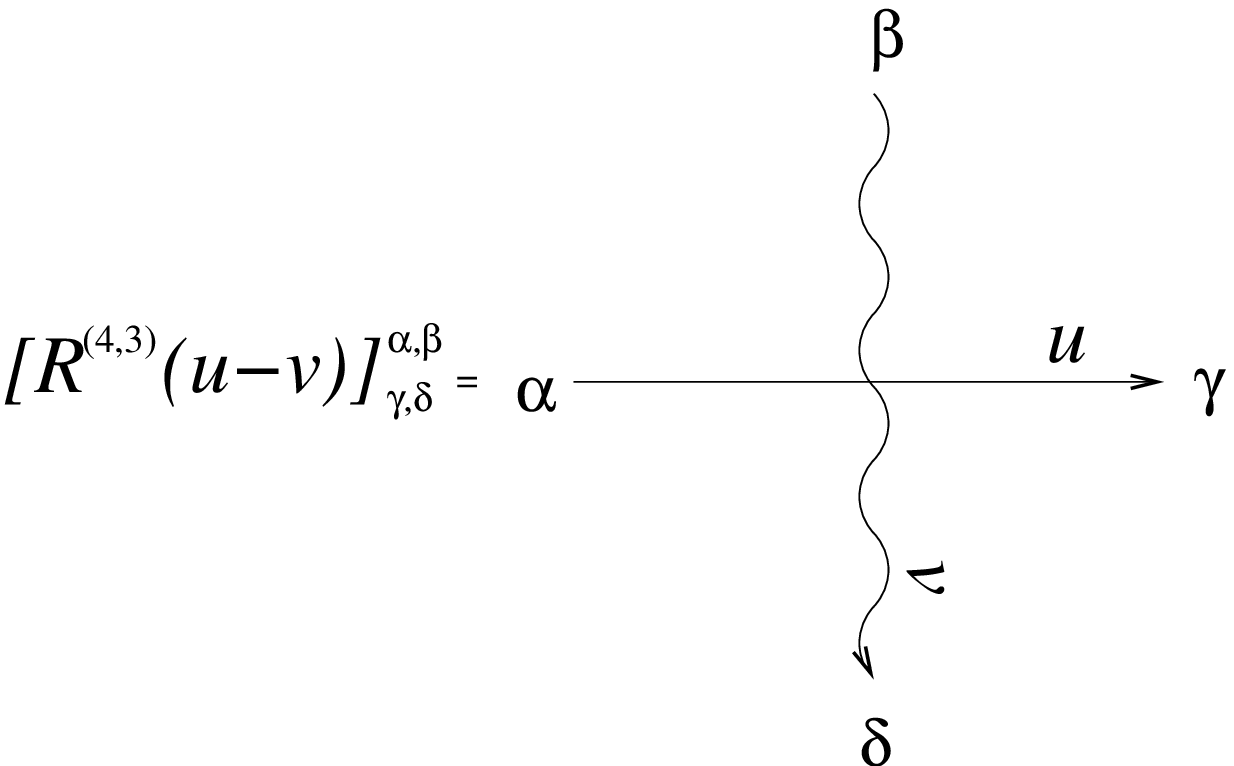}
\hspace*{2cm}
\includegraphics[scale=0.5]{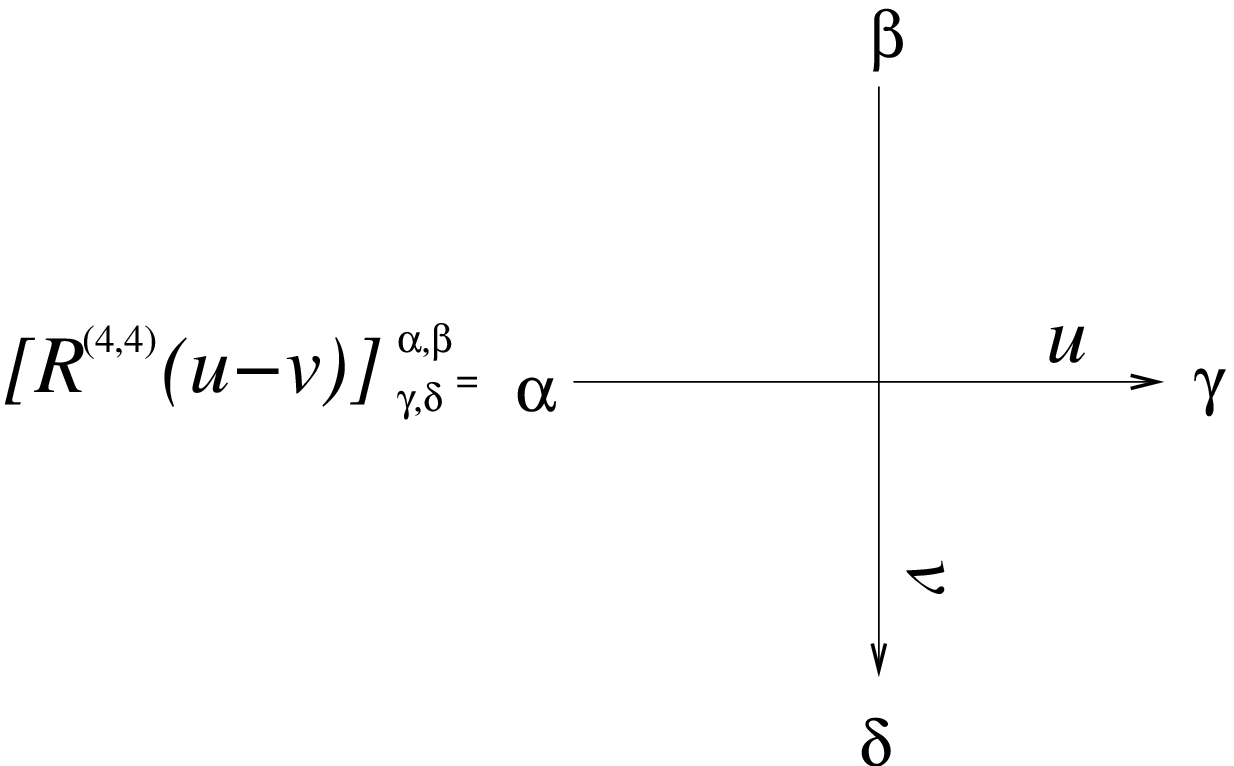}
\end{center}

The two lines symbolize the two spaces intertwined by $R_{i,j}$. Each line
carries a direction indicated by an arrow; both the vertical and horizontal lines carry spectral parameters. The argument
of $R$ is given by the difference between the right and the left ``incoming''
parameters. The replacement $R\to\ov R$ means flipping the arrow on the
vertical bond. 

The YBE eq. \refeq{ybedef} in graphical language reads:
\begin{center}
\includegraphics[scale=0.5]{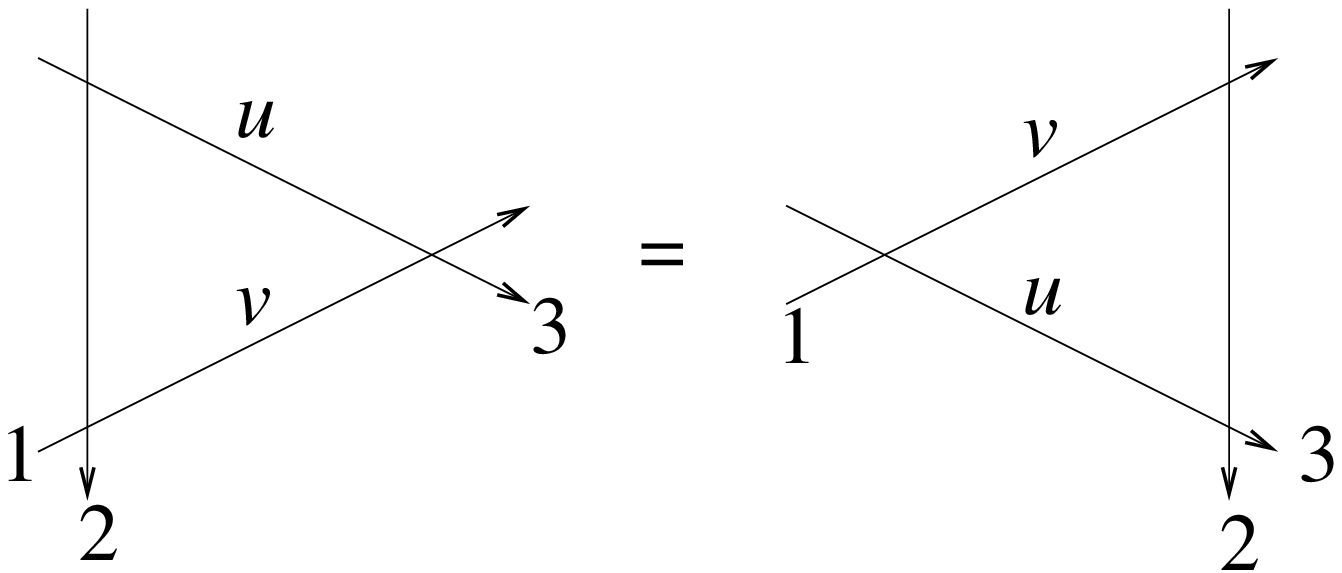}
\end{center}
``Other'' YBEs are obtained by flipping arrows (that means replacing
$R\to\ov R$) and/or substituting straight by wavy lines (that is, changing the
dimension in one of the spaces).
 
``Unitarity'' is a further property of the $R$-matrices. 
\be
\l[ R^{(d,d')}(u)\r]^{\beta,\g}_{\delta,\alpha}\,\l[
R^{(d',d)}(-u)\r]^{\alpha,\delta}_{\g',\beta'}=\delta^{\beta}_{\beta'}\,\delta^{\g}_{\g'}\label{uni}\;
,
\ee
and the same for $\ov R$. The unitarity property fixes normalization
constants of the $R$-matrices. In the following, we will speak of
``normalized'' $R$-matrices when they satisfy eq. \refeq{uni}; non-normalized
$R$-matrices differ from those by constant pre-factors, but still fulfill the
YBE. A direct verification of eq. \refeq{uni} for $d=d'=3$
($d=d'=4$) is done by using the projection properties
\be
e^a_b\,e^c_d&=&\delta^a_d\,e^c_b\;,\,\check P_j\,\check P_k=\delta_{j,k}\check
P_k \label{pro2}\; ,
\ee
or by using the YBE with standard initial conditions. Furthermore, for $d=3$, $d'=4$ in eq. \refeq{uni}, one should employ
\be
E^\alpha_\beta\,E^\beta_\delta (-1)^{p[\beta](p[\alpha]+p[\delta])}
=-(\alpha+2)(-1)^{p[\alpha]p[\delta]}E^{\alpha}_\delta\nn\;.
\ee
In order to construct a lattice model, one embeds $m_a^b$, $e_a^b$ into
End$\l[ V^{(3)}\otimes\l(V^{(4)}\r)^{\otimes L}\r]$, such that $e_a^b$ acts
non-trivially only on the lattice site $0$. Therefore consider the graded tensor product of two operators $v,w$:
\be
v_b^a\otimes_sw_d^c=(-1)^{(p[a]+p[b])p[d]} v_b^a\otimes w_d^c\nn\; ,
\ee 
where $v,w$ stand for $e,m$. The operator of unity in three-
(four-)dimensional space is $I_3=e_c^c$ ($I_4=m_c^c$). Following \cite{goe98}, define
\be
\l[e_0\r]^a_b&:=&e_b^a\otimes_sI_4^{\otimes_sL}\nn\\
&=&(-1)^{(p[a]+p[b])\sum_{k=1}^L p[c_k]}
e_b^a\otimes m^{c_{1}}_{c_{1}}\otimes \cdots\otimes m^{c_L}_{c_L}\nn\\
\l[m_j\r]^a_b&:=&I_3\otimes_sI_4^{\otimes_s(j-1)}\otimes_s m_b^a\otimes_sI_4^{\otimes_s(L-j)}\nn\\
&=&(-1)^{(p[a]+p[b])\sum_{k=j+1}^L p[c_k]}I_3\otimes I_4^{\otimes(j-1)}
m_b^a\otimes m^{c_{1}}_{c_{1}}\otimes \cdots\otimes m^{c_L}_{c_L}\nn\; ,
\ee
with $j=1,\ldots,L$. Then
\numparts
\label{subeq1}
\be
\l[e_0\r]^a_b\,\l[e_0\r]^c_d&=&\delta^a_d\,\l[e_0\r]^c_b\label{projprop}\\
\l[e_0\r]^a_b\,\l[m_k\r]^c_d&=&(-1)^{(p[a]+p[b])(p[c]+p[d])}[m_k]^c_d\,[e_0]^a_b\label{comprop}\; .
\ee
\endnumparts
Analogous relations hold between $m_j$, $m_k$. 

Principally, at this point one could derive the Hamiltonian. However, it is
more convenient to find a fermionic representation of the $R$ matrices in order to use the more
familiar language of fermionic field operators $c^\dagger _{\tau,j}, c_{\tau,j}$,
acting on the spin directions $\tau=\uparrow, \downarrow$ and on the lattice site $j$. This is done by employing the
technique of G\"ohmann \cite{goe98,goe00}, which consists in
identifying the $\l[m_j\r]^a_b$, $\l[e_0\r]_b^a$ with certain combinations of fermionic
operators. 

The entries $\l[X_j\r]^a_b$ of the matrix 
\be
X_j&=&\l(\begin{array}{cccc}
          n_{j\downarrow}n_{j\uparrow} &
          n_{j\downarrow}c^\dagger _{j\uparrow} & c^\dagger _{j\,\downarrow}n_{j\uparrow}
          & c^\dagger _{j\down}c^\dagger _{j\up}\\
          n_{j\down}c_{j\up} & n_{j\down}(1-n_{j\up}) &
          -c^\dagger _{j\down}c_{j\up} & -c^\dagger _{j\down}(1-n_{j\up})\\
          c_{j\down}n_{j\up}&c_{j\down}c^\dagger _{j\up}&(1-n_{j\down})n_{j\up}&(1-n_{j\down})c^\dagger _{j\up}\\
          -c_{j\down}c_{j\up}
          &-c_{j\down}(1-n_{j\up})&(1-n_{j\down})c_{j\up}&(1-n_{j\down})(1-n_{j\up})
       \end{array}\r)\label{tildeX}
\ee
satisfy projection and commutation properties formally identical to
eqs. \refeq{projprop}, \refeq{comprop} with grading $p[1]=p[4]=0$,
$p[2]=p[3]=1$ in accordance with
eq. \refeq{pardef} for $d=4$. This is the only constraint on
$\l[m_j\r]_b^a$, so that we identify $\l[X_j\r]^a_b\equiv \l[m_j\r]^a_b$. The
whole set \refeq{fourdimset} reads in fermionic language:
\be
\fl\begin{array}{ll}
\l[E_j\r]_3^3=\al+2 -\l(n_{j\down}+n_{j\up}\r)& \\
\l[E_j\r]_1^1=  n_{j\down}-1&\l[E_j\r]_2^2= n_{j\up}-1\\
\l[E_j\r]^1_2=-c^\dagger _{j\up} c_{j\down}& \l[E_j\r]^2_1=-c^\dagger _{j\down} c_{j\up}\nn\\
\l[E_j\r]^1_3=-\sqrt{\al}\,n_{j\up} c_{j\down}-\sqrt{\al+1}\,(1-n_{j\up})c_{j\down}&
\l[E_j\r]^3_1=-\sqrt{\al}\,n_{j\up} c^\dagger _{j\down}-\sqrt{\al+1}\,(1-n_{j\up})c^\dagger _{j\down}\\
\l[E_j\r]^2_3=\sqrt{\al}\,n_{j\down} c_{j\up}-\sqrt{\al+1}\,(1-n_{j\down})c_{j\up}&
\l[E_j\r]^3_2=\sqrt{\al}\,n_{j\down} c^\dagger _{j\up}-\sqrt{\al+1}\,(1-n_{j\down})c^\dagger _{j\up}
\end{array}
\label{erest}
\ee
The even sub-algebras are manifest: $E_3^3$ is the u(1)-generator, and
$E_{1,2}^{1,2}$ are the su(2) generators.  
The fermionization of the three-dimensional $e_b^a$ is done
with the matrix $Y$, resulting from $X$, eq. \refeq{tildeX} by deleting the first row
and column,
\be
Y&=& \l(\begin{array}{cc|c}
          n_{d,\down}(1-n_{d,\up}) &
          -d^\dagger _{\down}d_{\up} & -d_{\down}^\dagger (1-n_{d,\up})\\
         d_{\down}d^\dagger_{\up}&(1-n_{d,\down})n_{d,\up}&(1-n_{d,\down})d^\dagger_{\up}\\
          \hline -d_{\down}(1-n_{d,\up})&(1-n_{d,\down})d_{\up}&(1-n_{d,\down})(1-n_{d,\up})\\
       \end{array}\r)\label{Y}\; .
\ee
We slightly modified the notation, replacing $c^\dagger,c,n$ by $d^\dagger, d, n_d$. Horizontal and vertical bars separate fermionic and bosonic sectors. The
boxes on the diagonal of $Y$ contain the generators of su(2), u(1). Set $\l[e_0\r]_a^b=Y_a^b$, such that eqs. \refeq{projprop}, \refeq{comprop}
hold with grading $\l\{1,1,0\r\}$. 

The monodromy matrices
\be
T(u)&=&R^{(4,4)}_{a,L}(u)R^{(4,4)}_{a,L-1}(u)\ldots
R^{(4,4)}_{a,1}(u)R^{(4,3)}_{a,0}(u+\rmi u_0)\label{mondrom1}\\
\ov T(u)&=&\ov R^{(4,4)}_{a,L}(-u)\ov R^{(4,4)}_{a,L-1}(-u)\ldots \ov
R^{(4,4)}_{a,1}(-u)\ov R^{(4,3)}_{a,0}(-u+\rmi u_0)\nn
\ee
consist of sequences of $R$ matrices, multiplied in (horizontal) auxiliary
space. Note the shift by $\rmi u_0$ on the zeroth lattice site, where the dimension
of the (vertical) quantum space is reduced by one. This site shall be denoted as ''impurity
site''. The shift is done by $\rmi u_0\in \mathds C$, for reasons which will
become clear later. Graphically, $T(u)$ is depicted as
\begin{center}
\includegraphics[scale=0.5]{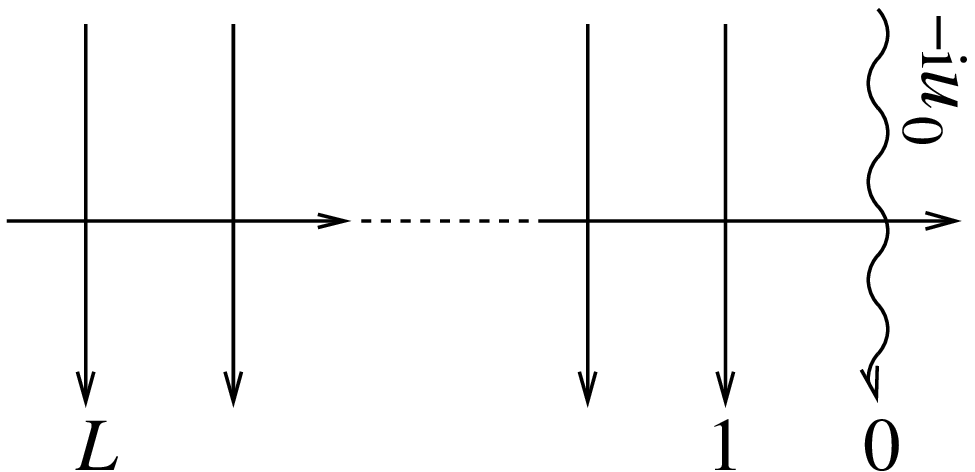}
\end{center}
The super-trace is called transfer matrix
\be
\tau(u)= \str_a T(u) \; , \qquad \ov \tau(u)=\str_a \ov T(u)\label{bartaudef}\\
\ln \l[\tau\ov\tau\r](u)= \ln
\l[\tau\ov\tau\r](0)+u\underbrace{\l[\tau^{-1}(0)\tau'(u)+\ov\tau^{-1}(0)\ov\tau'(u)\r]_{u=0}}_{\displaystyle{=:{\rm
    const.
    } H}}+\Or(u^2)\label{defham}\; .
\ee
In the last line, the Hamiltonian was defined as the logarithmic derivative of the
two transfer matrices at zero spectral parameter. By scaling $u$, one is free
to multiply the Hamiltonian by a constant factor.

Before evaluating eq. \refeq{defham}, let us shortly comment on the case of a
homogeneous model without impurity. We denote the
corresponding quantities with a subscript $h$. This model has been
extensively studied in \cite{bra95,bar95}. Assuming periodic boundary conditions $1\equiv
L+1$, $\tau_h(0)$ ($\ov\tau_h(0)$)
is the right (left)
shift operator, and  
\be
\ln
\tau_h(0)=\rmi P=-\ln \ov \tau_h(0)\label{phost}\; ,
\ee
where $P$ is the generator of translations to the right. The derivative with
respect to $u$ in eq. \refeq{defham} yields a sum of
$L$ terms, each one corresponding to $R_{j,j+1}'(0)$, $j=1,\ldots, L$. For the
ease of notation, let us follow the graphical depiction of \cite{kldiss}. The
following figure shows the $j$th term of
$\l[\tau_h^{-1}(0)\tau_h'(u)\r]_{u=0}$: 
\begin{center}
\includegraphics[scale=0.7]{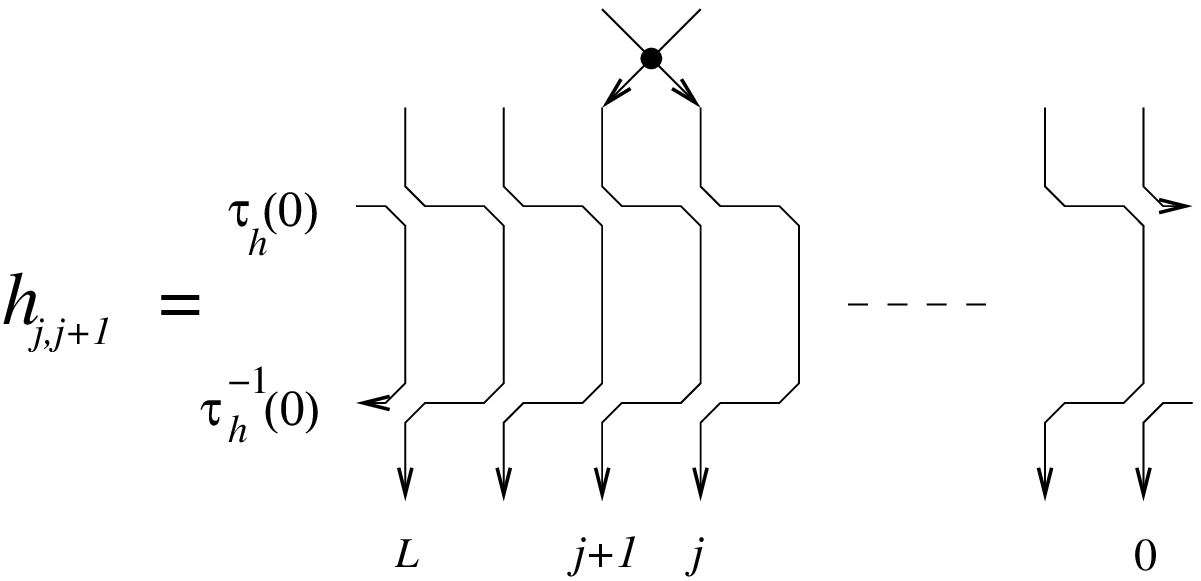}
\end{center}
The vertex
with a dot denotes $R_{j,j+1}'(0)$. For $u=0$, the
vertices decouple and taking the trace over a row yields the right shift
operator $\tau_h(0)$. Thus
\be
H_h&=& \sum_{j=1}^L h_{j,j+1}\label{loccon}\\
h_{j,j+1}&=& (\alpha+1)D\,\frac{\d}{\d u} \ln
\l[R^{(4,4)}(u)\r]_{j,j+1; u=0}\nn
\ee
$H_h$ is scaled by $D(\al+1)$, $D$ is a bandwidth parameter whose
significance will become clear later. Using explicit expressions for
$R_{j,j+1}^{(4,4)}$ from \cite{bra95} one confirms the expression for $h_{j,j+1}$
given in \cite{bar95}:
\be
\fl h_{j,j+1}=(\al+1)D \l(\frac{2}{\alpha} (\check P_1)_{j,j+1}
-\frac{2}{\alpha+1} (\check P_3)_{j,j+1}\r)\nn\\
\fl\qquad\;\;=-D\,\sum_\tau (c_{j,\tau}^\dagger c_{j+1,\tau} + c^\dagger _{j+1,\tau}c_{j,\tau})
\te^{-\frac{\eta}{2}(n_{j,\ov\tau}+n_{j+1,\ov\tau})}\nn\\
\lo+U\l(n_{j,\up}n_{j,\down}+n_{j+1,\up}n_{j+1,\down}\r)+t_p\l(c^\dagger _{j+1,\up}
c^\dagger _{j+1,\down}c_{j,\up}c_{j,\down}+c^\dagger _{j,\up}c^\dagger _{j,\down}c_{j+1,\up}c_{j+1,\down}\r)\nn\\
\lo  +D(n_j+n_{j+1}) - 2D\;,\label{hamhost}\\
\fl U=\frac{D}{\alpha}=t_p\;
,\;\te^{-\eta}=\frac{\alpha+1}{\alpha}\,;\; \ov \tau=-\tau\nn \; .
\ee
One can show by a canonical transformation \cite{bar95} that the limit $\al\to
0$ leads to the $t-J$-model. On the other hand, for $\al\gg 1$, the leading order are free fermions:
\be
\fl H_h/D=2 \sum_{j=1}^Ln_j-2-\sum_{j=1}^L\sum_\tau\l(c^\dagger_{j,\tau}
c_{j+1,\tau}+c^\dagger_{j+1,\tau} c_{j,\tau}\r)\l[1+\frac{1}{2\al}(n_{j,\bar\tau}+n_{j+1,\bar\tau})\r]\nn\\
\fl \qquad+\frac{2}{\al}\sum_{j=1}^Ln_{j,\up}n_{j,\down} +\frac
  1\al\sum_{j=1}^L\l(c^\dagger_{j+1,\up}c^\dagger_{j+1,\down} c_{j,\up} c_{j,\down}+c^\dagger_{j,\up}c^\dagger_{j,\down} c_{j+1,\up} c_{j+1,\down}\r)+\Or\l(D/\al^{3/2}\r)\label{hhall}.
\ee
Since in this work, we aim at realizing a free host, our interest is in the 
limit $\al\gg 1$.

Due to the insertion of $R^{(4,3)}_{a,0}$, $H$ receives an impurity contribution $H_i$. It
can be derived graphically. First observe that eq. \refeq{phost} no longer holds; but
due to unitarity \refeq{uni}, one still has $\tau(0)= \ov \tau^{-1}(0)$.
Then $\tau(0)$ is depicted as:
\begin{center}
\includegraphics[scale=0.5]{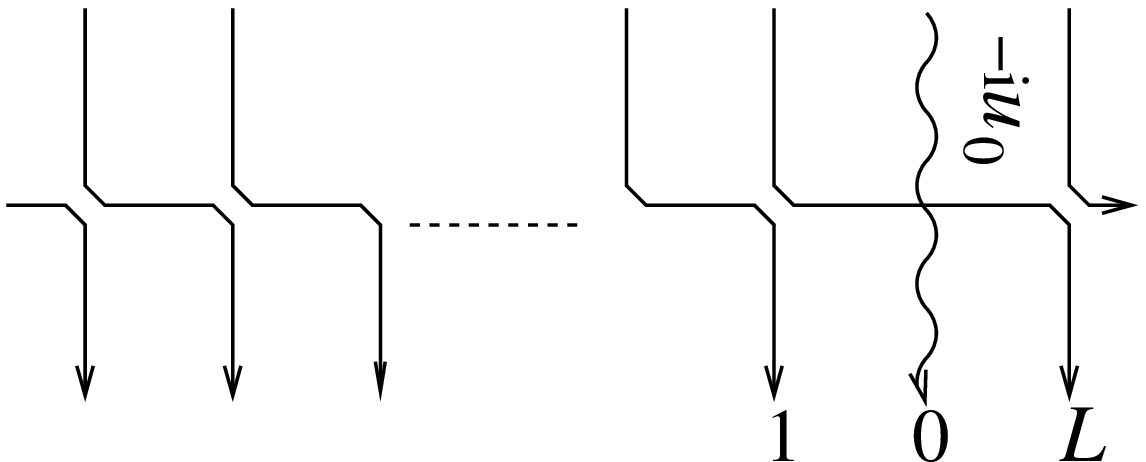}
\end{center}
In comparison with the free
host, the changes induced by the impurity  stemming from $\ln \tau' (0)$ correspond to the graphs:
\begin{center}
\includegraphics[scale=0.5]{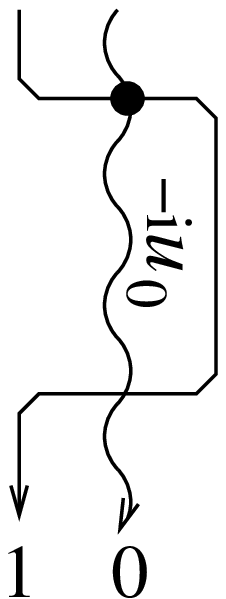}
\hspace*{4cm}
\includegraphics[scale=0.5]{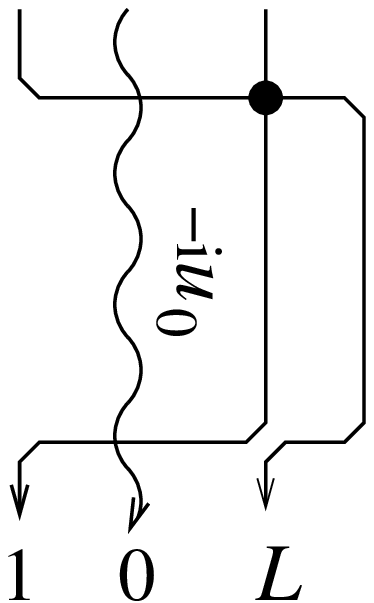}
\end{center}
A vertex with a dot symbolizes the derivative with
respect to the spectral parameter. 
The first term, 
\be
 R^{(3,4)}_{1,0}(-\rmi
u_0)\l[R^{(4,3)}_{1,0}\r]'(\rmi u_0)=R^{-1\;(4,3)}_{1,0}(\rmi u_0)\,\l[R^{(4,3)}_{1,0}\r]'(\rmi u_0)\nn\;,
\ee
couples the impurity to the left neighboring site. The second term, 
\be
R_{1,0}^{(3,4)}(-\rmi u_0)
 \frac{h_{L,1}}{D(\alpha+1)}R_{1,0}^{(4,3)}(\rmi u_0)=R_{1,0}^{-1\;(4,3)}(\rmi u_0)\,\frac{h_{L,1}}{D(\al+1)}\,R_{1,0}^{(4,3)}(\rmi u_0)\nn\;,
\ee
is a three site
coupling. Analogous terms, with $L$ and $1$ interchanged, are provided by $\ln
\ov\tau'(0)$. The inverse matrix $\l[R^{(4,3)}\r]^{-1}$ is found by
eq. \refeq{uni}. 
In analogy to the host Hamiltonian, the
spectral parameter $u$ is scaled by $D(\al+1)$, where $D$ is a
bandwidth parameter. Using the fermionization technique, we calculate $H_i$. It contains terms which are symmetric and others which are antisymmetric under $(L\leftrightarrow 1)$ exchange. In the combined thermodynamic and continuum limit, the latter do not interact with the impurity \cite{zvy97}, so we do not consider them in the following. Then
\be
\fl H_i=-2 DJ_\al (1-n_d)\l(\al+F_{1,L}\r)\mathcal{P}_d\nn\\
\lo - DJ_\al\,\sqrt\al\,\mathcal{P}_d\,\sum_\tau
\l[d^\dagger_\tau(c_{L,\tau}+c_{1,\tau})-d_\tau(c^\dagger_{L,\tau}+c^\dagger_{1,\tau})\r]\mathcal{P}_d\nn\\
\lo  + DJ_\al \sum_\tau
n_{d,\tau}\l(c^\dagger_{L,\tau}c_{1,\tau}+c^\dagger_{1,\tau}c_{L,\tau} \r)\mathcal{P}_d\nn\\
\lo  + DJ_\al\sum_\tau d^\dagger_\tau d_{\bar\tau}
\l(c^\dagger_{L,\bar\tau}c_{1,\tau}+c^\dagger_{1,\bar\tau}c_{L,\tau}\r)\mathcal{P}_d+\Or\l(DJ_\al/\al^{1/2}\r)\label{impham}\;,
\ee
where the projector $\mathcal{P}_d:=1-n_{d,\up}n_{d,\down}$ projects onto non-doubly occupied states on the impurity site. Furthermore, 
\be
 F_{1,L}&=&2-2\hat h-c^\dagger_{1,\up} c_{L,\up}-c^\dagger_{1,\down}
 c_{L,\down}-c^\dagger_{L,\up} c_{1,\up}-c^\dagger_{L,\down} c_{1,\down}\nn\\
J_{\al}&=&\frac{2\al}{v_0^2+\al^2}>0\;,v_0:=u_0/2\label{defjal}
\ee
with the most interesting range of the coupling constant
\be
\al^{-2}\lesssim J_\al\lesssim \al^{-1}\label{jalrange}\; .
\ee
In \ref{appa}, an
alternative fermionization is used, resulting in essentially the same
Hamiltonian.  

Finally, one includes external fields $\mu$, $h$, by
\be
\fl H_{ex}=\frac{h}{2}\l[\sum_{j=1}^L\l(n_{j,\up}-n_{j,\down}\r)+\l(n_{d,\up}-n_{d,\down}\r)\r]-\mu\l[\sum_{j=1}^L
n_j+n_{d}\mathcal{P}_d\r]\label{hamex}\; .
\ee
Eqs. \refeq{hhall}, \refeq{impham}, \refeq{hamex} define the entire
Hamiltonian of the impurity model in the limit of an asymptotically free host. In \ref{appb}, it is shown that
$H_h+H_i$ displays gl(2$|$1) symmetry. $H_{ex}$ breaks this symmetry, but
preserves integrability. 

The Kondo limit can be performed by a canonical transformation which
eliminates transitions between single and zero occupation of the impurity site. It
is conveniently done in Fourier space:
\be
c_{j,\tau}^\dagger&=&\frac{1}{\sqrt{L}}\sum_{k=-\pi}^{\pi} c_{k,\tau}^\dagger \,\te^{\rmi kj} \nn\\
H_h&=& D\l\{\sum_{k}\sum_\tau
\epsilon_kc^\dagger_{k,\tau}c_{k,\tau}-2\r.\nn\\
& &+\frac{2}{L\,\al}\sum_{Q,q,q'}
\l[\sum_\tau \cos \frac{Q}{2} \cos\l(q+\frac{Q}{2}\r)c^\dagger_{q+Q,\tau}
c_{q,\tau}
c^\dagger_{q'-Q,\bar\tau}c_{q',\bar\tau}\r.\nn\\
& &+\l.\l.c^\dagger_{q,\up}c_{q+Q,\up}c^\dagger_{q'-Q,\down}
c_{q',\down}
-\cos(q+q')c^\dagger_{q+Q,\up}c_{q,\up}c^\dagger_{q'-Q,\down}c_{q',\down}\r]\r\}\label{hhfouop}\\
\epsilon_k&=&2(\cos k+1)-\mu/D\label{hfffou}\\
H_i&=& \l\{2DJ_\al(n_{d}-1)(\al+F_{1,L})-\mu n_d\r\}\mathcal{P}_d\nn\\
& &+J_\al(\al D)^{1/2}\P_d\sum_{k,\tau}(M_kd^\dagger_\tau
c_{k,\tau}+M_k^*c^\dagger_{k,\tau}d_\tau )\P_d\nn\\
& &
+J_\al\sum_{\tau,k,k'} N_{k,k'}\l[n_{d,\tau}c^\dagger_{k,\tau}c_{k',\tau}
+d^\dagger_\tau d_{\bar \tau} c^\dagger_{k,\bar\tau} c_{k',\tau}\r]\P_d \label{fouham}\\
M_k&=& -\,\frac{1}{\sqrt l}\l(1+\te^{\rmi k}\r)\;,\,N_{k,k'}=\frac{1}{l}\l(\te^{-\rmi k'}+\te^{\rmi k}\r)\nn\; .
\ee
Here $l=L/D$ is the constant length of the chain, and $D^{-1}$ plays the role of a lattice
constant. The canonical transformation is generated by an operator $A$, which yields a
transformed Hamiltonian $H_{eff}=\exp(A)H\exp(-A)$ not containing any
hybridization between impurity and host in the leading order ${\rm
O}(J_\al)$. One verifies that 
\be
A=J_\al\sqrt{\frac{\al}{D}}\,\P_d\,\sum_{k,\tau}
\frac{1}{\epsilon_d-\epsilon_k}(M_kd^\dagger_\tau c_{k,\tau}-M^*_kc^\dagger_{k,\tau}
d_\tau)\P_d\label{defA}\; ,
\ee
where $\epsilon_d:=2J_\al \al n_d-\mu/D$ has been defined. $H_{eff}$ contains terms $\Or(\al J_\al^2)$. Given the restriction
\refeq{jalrange}, these terms can be neglected. After the transformation, the
excitation spectrum of the impurity site contains only the contribution for single
occupation. The contribution for
non-occupation in the impurity operator is energetically suppressed in the strong coupling limit: In the language of
renormalization theory, $v_0$ drives the impurity Hamiltonian to a strong
coupling fixed point at low temperatures. This will be demonstrated in the next
section. Thus we do not
consider the contribution from zero occupation to the Hamiltonian in the ongoing.  

To perform the scaling limit, the fermionic spectrum is linearized around
incommensurate Fermi points $\pm k_F$ avoiding Umklapp scattering. The
linearization gives rise to right (left) moving particles $R$ ($L$). To avoid divergences due to the
unbounded linear spectrum, operator products are normal ordered:
\be
{\rm :} c^\dagger_{k,\nu,\tau} c_{k,\nu',\tau'}{\rm :}=c^\dagger_{k,\nu,\tau}
c_{k,\nu',\tau'}-\langle
c^\dagger_{k,\nu,\tau}c_{k,\nu',\tau'}\rangle_0\nn\;,
\ee
where $\nu\in\{R,L\}$. The continuous description is achieved by introducing
field operators \cite{goeh03}:
\be
c^\dagger_{k,\nu,\tau}&=&\frac{1}{\sqrt L} \sum_{n=1}^L\te^{\rmi k_\nu
  n}c^\dagger_{n,\nu,\tau}\nn\\
&=&\sqrt{\frac{D}{L}} \sum_{n=1}^L \te^{\rmi k_\nu D
  \frac{n}{D}}c^\dagger_{n,\nu,\tau}\sqrt{D}\frac{1}{D}\nn\\
&=& \frac{1}{\sqrt l} \int_0^l \te^{\rmi q_{k,\nu} x} \psi^\dagger_{\nu,\tau}(x)\,\d x\nn\;,
\ee
where $x=n/D$,
$\psi^\dagger_{\nu,\tau}(x)=\lim_{D\to \infty}\sqrt{D}c^\dagger_{n,\nu,\tau}$,
$q_k=k\cdot D$. $\psi^\dagger$, $\psi$ are now fermionic field operators with$
\l\{\psi_{\nu,\tau}(x),\psi^\dagger_{\nu',\tau'}(x')\r\}=\delta(x-x')\delta_{\nu,\nu'}\delta_{\tau,\tau'}$.
Again normal
ordering is imposed,
\be
{\rm :}\psi^\dagger_{\nu,\tau}(x)\psi_{\nu',\tau'}(x){\rm :}=\lim_{\e\to 0}\l[\psi^\dagger_{\nu,\tau}(x+\e)\psi_{\nu',\tau'}(x)-\langle
\psi^\dagger_{\nu,\tau}(x+\e)\psi_{\nu',\tau'}(x)\rangle_0\r]\nn\; ,
\ee
where $\langle\cdots\rangle_0$ is the expectation value in the ground
state. Let us summarize the external fields again in an operator
$H_{ex}$. Then 
\numparts
\label{subeq2}
\be
\fl \frac{H_h}{2D}=\int_0^l\sum_{\nu,\nu',\tau}\l[ \delta_{\nu,\nu'}(\cos
k_F+1)\,{\rm :}n_\nu{\rm :}-\delta_{\nu,\nu'}\mathds 1_\nu+\rmi\nu \frac{\sin k_F}{D}{\rm :}\psi^\dagger_{\tau,\nu}(x)\frac{\d}{\d
  x}\psi_{\tau,\nu'}(x){\rm :}\r] \,\d x\;,\label{hostcon}\\
\fl H_i= 2\cos k_F J\int_0^l \delta(x)\sum_{\nu,\nu',\tau}\l[\delta_{\nu,\nu'}n_{d,\tau}\P_d{\rm :}n_{\nu,\tau}(x){\rm :}+d^\dagger_\tau
d_{\bar \tau} {\rm :}\psi^\dagger_{\nu,\bar \tau}(x)\, \psi_{\nu',\tau}(x){\rm :}\r]\d x\;,\label{impcon}\\
\fl H_{ex}= \int_0^l-\mu \l[ n(x)+\delta(x)
n_{d}\P_d\r]+\frac{h}{2} \l[\delta(x)(n_{d,\up}-n_{d,\down})+(n_\up(x)-n_\down(x) ) \r]\d x\; .\label{excon}
\ee
\endnumparts
The occupation number operators are $n_{\tau,\nu}={\rm :}\psi^\dagger_{\tau,\nu}\psi_{\tau,\nu}{\rm :}$, $n_\nu=\sum_\tau
n_{\tau,\nu}$, $n_\tau=\sum_\nu n_{\tau,\nu}$. 

As far as the terms \refeq{hostcon}, \refeq{impcon} are concerned, one may pass to a Weyl basis by the
canonical transformation
\be
\fl \phi_{\pm,\tau}(x)= \frac{1}{\sqrt2}\l[\psi_{L,\tau}(x)\pm
\psi_{R,\tau}(-x)\r]\;,\qquad\l\{\phi_{\nu,\tau}(x),\phi^\dagger_{\nu',\tau'}(x')\r\}&=&\delta(x-x')\delta_{\nu,\nu'}\delta_{\tau,\tau'}\nn\; .
\ee
Interaction terms in the host are non-local in the $\phi_\pm(x)$; however, as
will be shown in the next section these are accounted for by a redefinition of the
Fermi velocity $v_F$, $\sin k_F=v_F\to \tilde v_F=v_F(1+\Or(1/\al))$. The Weyl
basis demonstrates that the impurity
couples only with one of the two host channels. The Hamiltonian density thus reads:
\be
\fl \mathcal H_h= 2\sum_{\tau,\nu=\pm}\l[\rmi \tilde v_F{\rm :}\phi^\dagger_{\nu,\tau}(x)
\frac{\d}{\d x} \phi_{\nu,\tau}(x){\rm :}+D(\cos k_F+1){\rm :}n_{\nu}(x){\rm :} -D\mathds 1_\nu\r]\label{defhhlindis}\\
\fl\mathcal H_i= 4 J \cos k_F\sum_\tau \delta(x)\l[{\rm :} \phi^\dagger_{+,\tau}(x) \phi_{+,\tau}(x){\rm :}
n_{d,\tau}\P_d+ {\rm :}\phi^\dagger_{+,\tau}(x) \phi_{+,\bar \tau}(x){\rm :}
d^\dagger_{\bar\tau}d_{\tau}\r]\nn\\
\fl\mathcal H_{ex}=- \mu \l[n(x)+\delta(x)n_{d}\r]+\frac{h}{2} \l[\delta(x)(n_{d,\up}-n_{d,\down})+ n_\up(x)-n_\down(x) \r]\nn\; .
\ee
The fermionic operators of
the impurity can be expressed in terms of spin operators with index
$i$,
\be
\sigma^z_i=n_{d,\up}-n_{d,\down}\,,\;\sigma^+_i=d^\dagger_\up
d_\down\,,\;\sigma^-_i=d^\dagger_\down d_\up\nn\; .
\ee
Then one directly recognizes that the impurity operator is su(2)-symmetric and can be completed to
the XXX-exchange operator,
\be
\mathcal
H_{i}&=&2 J\delta(x)\sum_{\tau,\tau'}{\rm :}\phi^\dagger_{+,\tau}(x)
{\bsigma}_{\tau,\tau'}\phi_{+,\tau'}(x){\rm :}{\bsigma}_i+2 J\delta(x)n_d\P_d{\rm :} n_+(x){\rm :}\label{hsd}\; ,
\ee
where $\bsigma=(\sigma^x,\sigma^y,\sigma^z)^T$ and $2\cos k_F J_\al=: J$ is defined. 
Eq. \refeq{hsd} constitutes the isotropic Kondo model.

\section{Calculation of the free energy}
\begin{figure}
\begin{center}
\includegraphics[scale=0.6]{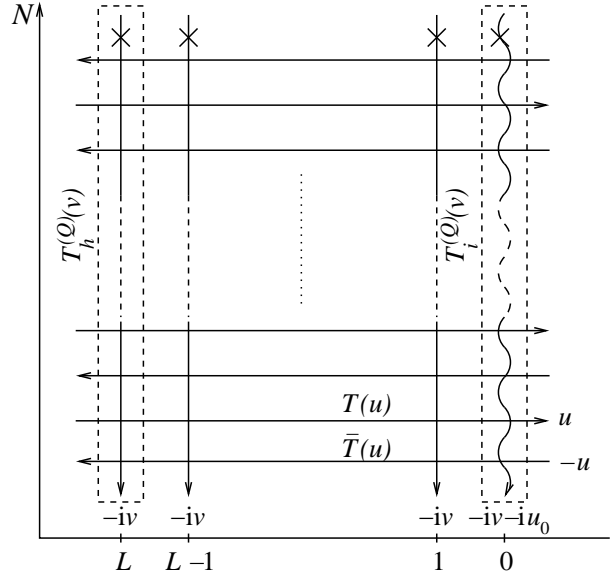}
\end{center}
\caption{Classical lattice representing the free energy of the impurity
  model. $L$ is the physical, $N$ the Trotter direction. The dimension in the
  impurity quantum space (wavy line) is reduced by one. The spectral parameter on the straight vertical lines is $0$ and on the wavy vertical line $-\rmi u_0$. For the further analysis it is convenient to introduce the auxiliary spectral parameter $-\rmi v$ on the vertical lines. Crosses stand for
  twisted boundary conditions, induced by external fields $h$ and $\mu$.}
\label{fig1}
\end{figure}

Taking account of eq. \refeq{defham}, 
\be
\fl \te^{-\beta H_h}= \lim_{N\to\infty}\l[ \bar \tau_h(u_N) \,\tau_h(u_N)\r]^{N/2}\; , \;
u_N=-\frac{\beta D(\alpha+1)}{N}\nn\\
\fl \te^{-\beta H}= \lim_{N\to\infty}\l[ \bar \tau(u_N)
\,\tau(u_N)\r]^{N/2}\te^{-\beta H_{ex}}\nn\\
\fl \te^{-\beta H_{ex}} = \prod_{j=1}^L \te^{-\beta
  [h/2(n_{j,\up}-n_{j,\down})-\mu
  n_j]}\,\te^{-\beta\l[h/2(n_{d,\up}-n_{d,\down})-\mu \sum_\tau
  n_{d,\tau}\r]}=:\te^{-\beta\sum_{j=1}^L h_{ex,j}}\te^{-\beta H_{ex,i}}\nn\,.
\ee
The even integer $N$ is referred to as
Trotter number and is the height of the fictitious underlying square
lattice, see fig. \refeq{fig1}. The impurity contribution to the free energy is 
\be
\fl f_i=-\lim_{ L\to\infty} \lim_{N\to\infty}\frac{1}{\beta}\l\{\ln \tr\l[ \l[\bar
\tau(u_N)\,\tau(u_N)\r]^{N/2}\te^{-\beta H_{ex}}\r]-\ln \tr\l[\l[\bar
\tau_h(u_N)\,\tau_h(u_N)\r]^{N/2}\te^{-\beta H_{ex,h}}\r] \r\}\nn ,
\ee
where $H_{ex,h}=\sum_{j=1}^L h_{ex,j}$. The crucial idea in calculating $f_i$ is to exchange $\tr$ and
$\str$ in the expression
\be
\fl \tr\l\{\l[ \ov \tau(u_N) \,\tau(u_N)\r]^{N/2}\te^{-\beta H_{ex}}\r\}= \tr\te^{-\beta H_{ex}} \prod_{k=1}^{N/2} \str_{a_{2k} a_{2k-1}}
\l[\ov R_{a_{2k} L}^{(4,4)}(-u_N) \ldots \ov R_{a_{2k}
  1}^{(4,4)}(-u_N)\r.\nn\\
\lo\times\l.\ov R_{a_{2k} 0}^{(4,3)}(-u_N+\rmi u_0)R^{(4,4)}_{a_{2k-1} L}(u_N)
\cdots R^{(4,4)}_{a_{2k-1} 1}(u_N)R^{(4,3)}_{a_{2k-1} 0}(u_N+\rmi u_0)\r]\nn\,.
\ee
This leads to
\be
\fl\str \prod_{j=1}^L \l[\tr_j\te^{-\beta h_{ex,j}}\prod_{k=1}^{N/2} \ov R^{(4,4)}_{a_{2k} j}(-u_N)
R^{(4,4)}_{a_{2k-1}j}(u_N)\r]\,\nn\\
\fl\qquad\times\l[\tr_0\te^{-\beta H_{ex,i}}\prod_{k=1}^{N/2} \ov
R^{(4,3)}_{a_{2k} 0}(-u_N+\rmi u_0)
R^{(4,3)}_{a_{2k-1}0}(u_N+\rmi u_0)\r] =:\str \l[\tau^{(Q)}_h(0)\r]^L\tau^{(Q)}_i(u_0)\nn\\
\fl\tau^{(Q)}_h(v):=\tr_j \,\te^{-\beta h_{ex,j}}\prod_{k=1}^{N/2} \ov
R^{(4,4)}_{a_{2k} j}(-u_N+\rmi v)
R^{(4,4)}_{a_{2k-1}j}(u_N+\rmi v)=:\tr_j T^{(Q)}_h(v) \label{defqtmh}\\
\fl \tau^{(Q)}_i(v):=\tr_0 \,\te^{-\beta H_{ex,i}}\prod_{k=1}^{N/2} \ov
R^{(3,4)}_{0,a_{2k}}(-u_N+\rmi v)
R^{(3,4)}_{0,a_{2k-1}j}(u_N+\rmi v)=:\tr_0 T^{(Q)}_i(v) \label{defqtmi}\; .
\ee 
Eqs. \refeq{defqtmh}, \refeq{defqtmi} define the Quantum Transfer Matrix (QTM)
$\tau^{(Q)}_h$ of the host and $\tau^{(Q)}_i$ of the impurity,
respectively. Note that the host matrix is independent of the lattice site
$j$. Each QTM is the trace over the auxiliary space of a Quantum Monodromy Matrix $
T^{(Q)}$. The auxiliary space of $\tau^{(Q)}_h$ is four-dimensional, of
$\tau^{(Q)}_i$ three-dimensional. Fig. \ref{fig1} depicts this ``rotation''
from auxiliary space into quantum space. 
Due to eqs. \refeq{ybedef}, \refeq{ybebardef}, 
\be
\l[ \tau^{(Q)}_\nu(v),\tau^{(Q)}_{\nu'}(v')\r]&=&0\label{taucom}\,,
\ee
where the symbolical indices $\nu,\nu'$ may take values $h,i$. The auxiliary spectral parameter is essential for the diagonalization of
$\tau^{(Q)}$, the $u_N$ are inhomogeneities with alternating
signs. Especially, eq. \refeq{taucom} holds for $\nu\neq \nu'$: The impurity and
host QTM's share the same set of eigenvectors. The largest eigenvalue of
$\tau^{(Q)}_\nu$ is separated by a gap from the rest of the spectrum for any
$N$. The eigenstate $|\Phi_{\max}\rangle$ leading to
the largest eigenvalue $\L^{\max}_i(v)$ of
$\tau^{(Q)}_i(v)$ also leads to the largest eigenvalue $\L^{\max}_h(v)$ of
$\tau^{(Q)}_h(v)$. Although interesting, this is not essential: The dominant eigenstate $|\Phi_{\max}\rangle$  of the host matrix $\tau^{(Q)}_h(v)$ determines the ``correct'' eigenvalue of the impurity matrix $\tau_i^{(Q)}$: 
\numparts
\be
\label{subeq3}
\fl \ln\l[\str \l(\tau^{(Q)}_h(0)\r)^L
  \tau^{(Q)}_0(u_0)\r]\nn\\
\fl\qquad=\ln\l[(-1)^{p[\max]}\l(\L_h^{\max}(0)\r)^L\L_i^{\max}(u_0)+\sum_{k\neq\max}(-1)^{p[k]}\l(\L_h^{(k)}(0)\r)^L\L_i^{(k)}(u_0)\r]\label{gradev}\\
\fl\qquad\sim\ln\l[\l(\L_h^{\max}(0)\r)^L\L_i^{\max}(u_0)\r]+\sum_{k\neq\max}(-1)^{p[k]}\l(\frac{\L_h^{(k)}(0)}{\L_h^{\max}(0)}\r)^L\frac{\L_i^{(k)}(u_0)}{\L_i^{\max}(u_0)}\label{deffimpb}\;,
\ee
\endnumparts
which is an asymptotical expansion for large $L$. Generally, the eigenstate of the $k$th largest eigenvalue of
$\tau_h^{(Q)}$ does not lead to
the $k$th largest eigenvalue of $\tau_i^{(Q)}$. So with respect to
$\tau_i^{(Q)}$, $k$ does not label the eigenvalues according to their order. The supertrace requires to
include the parity of the projector on the eigenstate $k$. Note that $p[\max]=0$. In \cite{suz87} it is argued that the two limits $N\to\infty$, $L\to\infty$ are interchangeable. Then the thermodynamic limit $L\to \infty$ in eq. \refeq{deffimpb} is carried out just by keeping the largest eigenvalues $\L_{h,I}^{\max}$. 

One concludes that the impurity
and host contribution to the free energy per lattice site are given by
\be
f_i&=&-\lim_{N\to\infty}\frac{1}{\beta} \ln \Lambda_{i}^{\max}(u_0)\label{fimp}\\
f_h&=&-\lim_{N\to\infty}\frac{1}{\beta} \ln \Lambda_{h}^{\max}(0)\label{fh}
\ee
Eqs. \refeq{fimp}, \refeq{fh}
summarize the enormous advantage of considering the QTM: The calculation of
the free energy is reduced to the evaluation of a single eigenvalue. 

$f_h$ has already been calculated in this approach \cite{jue97,sak01}, the
result is given below. 
The diagonalization of $\tau^{(Q)}_h$ is done by
applying techniques of the nested Algebraic Bethe Ansatz (NABA)
\cite{goe02}, yielding for the eigenvalue $\L_i$ of the non-normalized QTM
$\tau_i^{(Q)}$: 
\be
\L_i(v)&=& \lambda_-(v)+\lambda_+(v)+\lambda_0(v)\label{ev}\\
\lambda_-(v)&=& \frac{q_-(v+\rmi)}{q_-(v)}\phi_+(v+\rmi
\al/2)\phi_-(v-\rmi\al/2-\rmi)\te^{\beta(\mu+h/2)}\nn\\
\lambda_+(v)&=& \frac{q_+(v-\rmi)}{q_+(v)}
\phi_-(v-\rmi\al/2)\phi_+(v+\rmi\al/2+\rmi)\te^{\beta(\mu-h/2)}\nn\\
\lambda_0(v)&=& \frac{q_-(v+\rmi)q_+(v-\rmi)}{q_-(v)q_+(v)} \phi_+(v-\rmi\al/2)\phi_-(v+\rmi\al/2)\nn\\
\nn\\
q_+(v)&=& \prod_{j=1}^M(v-v_j)\;, q_-(v)= \prod_{k=1}^{\widetilde M}(v-\tilde
v_k)\nn\; , \phi_{\pm}(v)=(v\pm \rmi u)^{N/2}\; .
\ee
An external magnetic field $h$ and a chemical potential $\mu$ have been
introduced. The roots - or particle solutions -  $\{v_j\}$, $\{\tilde v_k\}$ are determined by the analyticity of the
eigenvalue:
\numparts
\be
\fl\frac{\lambda_+(v_j)}{\lambda_0(v_j)}&=&\l.\frac{q_-(v)}{q_-(v+\rmi)}\,\frac{\phi_-(v-\rmi\al/2)\,\phi_+(v+\rmi+\rmi\al/2)}{\phi_+(v-\rmi\al/2)\,\phi_-(v+\rmi\al/2)}\,\te^{\beta(\mu-h/2)}\r|_{v=v_j}=-1\label{bae1}\\
\fl\frac{\lambda_-(\tilde v_k)}{\lambda_0(\tilde
  v_k)}&=&\l.\frac{q_+(v)}{q_+(v-\rmi)}\,\frac{\phi_+(v+\rmi\al/2)\,\phi_-(v-\rmi-\rmi\al/2)}{\phi_+(v-\rmi\al/2)\,\phi_-(v+\rmi\al/2)}\,\te^{\beta(\mu+h/2)}\r|_{v=\tilde v_k}=-1\label{bae2}\; .
\ee
\endnumparts
These Bethe-Ansatz equations are $M+\widetilde M$ many nonlinear coupled algebraic equations for the
unknown roots. Using analyticity properties, we represent the eigenvalue by a finite set of non-linear integral equations (NLIE). Within this approach, the Trotter limit $N\to\infty$ is carried out analytically. Consider the following combinations of $\lambda_{\pm,0}$, eq. \refeq{ev}: 
\be
\fl \frac{1}{\b(v)}:=
\frac{\la_+(v)}{\la_-(v)}\l(1+\frac{\la_0(v)}{\la_+(v)}\r)\nn\\
\lo=
\frac{q_+(v-\rmi)}{q_-(v+\rmi)}\,\frac{\phi_-(v-\rmi\al/2)\,\phi_+(v+\rmi+\rmi\al/2)}{\phi_+(v+\rmi\al/2)\,\phi_-(v-\rmi-\rmi\al/2)}\,\te^{-\beta h
  }\nn\\
\lo \;\times\;\underbrace{\frac{q_-(v)}{q_+(v)}\l(1+\frac{q_-(v+\rmi)}{q_-(v)}\,\frac{\phi_+(v-\rmi\al/2)\,\phi_-(v+\rmi\al/2)}{\phi_-(v-\rmi\al/2)\,\phi_+(v+\rmi+\rmi\al/2)}\te^{-\beta(\mu-h/2)}\r)}_{\displaystyle{\equiv \frac{q_-^{(h)}(v)}{\phi_-(v-\rmi\al/2)\,\phi_+(v+\rmi+\rmi\al/2)}}}\nn\\
\lo=\frac{1}{\phi_+(v+\rmi\al/2)\phi_-(v-\rmi-\rmi\al/2)}\,\frac{q_+(v-\rmi)}{q_-(v+\rmi)}\,q_-^{(h)}(v)\te^{-\beta
  h}\label{qhdef}\;,
\ee
where $q_-^{(h)}:=\prod_{j=1}^{N-M+\widetilde M}\l(v-\widetilde v_j^{(h)}\r)$, and $\widetilde v_j^{(h)}$ are called hole solutions. The polynomial $q_-^{(h)}$ has been identified by reasons of analyticity:
The zeroes of numerator and denominator cancel as far as the particle
solutions are concerned, the hole solutions rest as zeroes of the
numerator. The polynomials in the denominator are the same as the $\phi$-terms
in $\lambda_0/\lambda_+$. Along the same reasoning (or simply by taking the complex
conjugate and $h\to-h$), we find another function $\bb$:
\be
\frac{1}{\bb(v)}&:=
&\frac{\la_-(v)}{\la_+(v)}\l(1+\frac{\la_0(v)}{\la_-(v)}\r)\nn\\
&=&\frac{1}{\phi_-(v-\rmi\al/2)\phi_+(v+\rmi+\rmi\al/2)}\,\frac{q_-(v+\rmi)}{q_+(v-\rmi)}\,q_+^{(h)}(v)\te^{\beta
  h}\label{qqhdef}\; ,
\ee
where $q_+^{(h)}:=\prod_{j=1}^{N-\widetilde M+ M}\l(v-v_j^{(h)}\r)$.
A third function $\c$ is introduced,
\be
\fl\frac{1}{\c(v)}:=
\frac{\lambda_0(v)}{\lambda_+(v)\,\lambda_-(v)}\,\Lambda_i(v)\nn\\
\lo= \frac{\phi_+(v-\rmi\al/2)
  \,\phi_-(v+\rmi\al/2)\,\te^{-2\beta \mu}}{\phi_-(v-\rmi\al/2)\,\phi_+(v+\rmi+\rmi\al/2)\,\phi_+(v+\rmi\al/2)\,\phi_-(v-\rmi\al/2-\rmi)}\,\Lambda_i(v)
\label{defc}\; .
\ee
Consider $\B:=1+\b$, $\BB:=1+\bb$, $\C:=1+\c$,
\be
\B(v)&=& \frac{1}{\lambda_-(v)}\,\b(v)\,\Lambda_i(v)=
\frac{q_-(v)}{q_+(v-\rmi)\,q_-^{(h)}(v)}\,\Lambda_i(v)\,\te^{-\beta(\mu-h/2)}\nn\\
\BB(v)&=& \frac{1}{\lambda_+(v)}\,\bb(v)\,\Lambda_i(v)=
\frac{q_+(v)}{q_-(v+\rmi)\,q_+^{(h)}(v)}\,\Lambda_i(v)\,\te^{-\beta(\mu+h/2)}\nn\\
\C(v)&=&
\frac{1}{\b(v)\,\bb(v)}\,\c(v)=\frac{q_+^{(h)}(v)\,q_-^{(h)}(v)}{\phi_+(v-\rmi\al/2)\,\phi_-(v+\rmi\al/2)}\,\frac{1}{\L_i(v)}\,\te^{2\beta\mu}\nn\;.
\ee
In \ref{appb}, the unknown functions $q_{\pm}, q_{\pm}^{(h)},\L_i$ are
  expressed through the auxiliary functions by means of analyticity
  arguments for the largest eigenvalue. The result is a closed set of NLIE:
\numparts
\label{subeq9}
\be
\fl\ln \b(v)=\phi_\b(v+\rmi\delta) -[k_\b*\ln\BB](v+2\rmi\delta)-[k_\b*\ln\C](v+\rmi\delta)+\beta(\mu+h/2)\label{beqv2}\\
\fl\ln \bb(v)=\phi_\bb(v-\rmi\delta) -[k_\bb*\ln\B](v-2\rmi\delta)-[k_\bb*\ln\C](v-\rmi\delta)+\beta(\mu-h/2)\label{bbeqv2}\\
\fl\ln\c(v)=\phi_\c(v)
-[k_\b*\ln\BB](v+\rmi\delta)-[k_\bb*\ln\B](v-\rmi\delta)-[k_\c*\ln\C](v)+2\beta\mu\label{ceqv2}
\ee
\endnumparts
The contributions of the impurity and host to the free energy are given by:
\be
-\beta f_i&=&-\ln \c(u_0)-2\beta D(\al+1) J_\al+2\beta
\mu\label{ffromc}\\
-\beta f_h&=&\eta(0)+[\zeta*\ln \B](0)+[\ov\zeta*\ln \BB](0)+[(\zeta+\ov\zeta)*\ln \C](0)\label{evhub}
\ee
The inhomogeneities are
\be
\phi_\b(v)&=&-\,\frac{\beta D(\alpha+1)^2}{(v+\rmi\al/2)(v-\rmi\al/2-\rmi)}\;,\;\phi_\bb=\phi_\b^*\label{phib}\\
\phi_\c&=&\phi_\b+\phi_\bb\label{phic}\; .
\ee
The convolutions $[f*g](x):=\i f(x-y)g(y)\d y$ involve local kernels:
\be
k_\b(v)&=&\frac{1}{2\pi v(v-\rmi)}\, ,\; k_\bb=k_\b^*\,,\;k_\c=
k_\b+k_\bb\nn\\
2 \pi \zeta(v)&=& -\frac{\phi_\b(-v)}{D\beta(\al+1)}\, ,\; \eta(v)= 2\beta
D\frac{(\al+1)^2}{v^2+(\al+1)^2}\nn\; .
\ee
Following the treatment of the Hamiltonian in the preceding section,
we want to perform an asymptotic expansion of the free energy in the limit
$\al\gg 1$. The essential observation from the study
of the Hamiltonian is that after the canonical transformation, excitations of
the impurity stem exclusively from transitions between singly occupied
states. Furthermore, it was argued that $u_0$ has to be scaled such that zero
occupation is energetically suppressed. First consider the case where $u_0$ is
held fixed. By scaling $v\to\al v$, the algebraically decaying kernels shrink
to $\delta$-functions. The leading contribution thus solely stems from the
driving terms $\phi_{\b,\bb,\c}$. Within this
approximation, the auxiliary functions can be calculated explicitly:
\be
\fl \b(\al v)=\frac{a(\al v)}{1+\ov a(\al v)}\; ,\; \bb(\al v)=\frac{\ov a(\al v)}{1+a(\al v)}\;,\;
\c(\al v)=\frac{a(\al v)\ov a(\al v)}{1+a(\al v)+\ov a(\al v)}\label{auxlimff}\\
\fl a(v)=\exp\l[\phi_\b(v+\rmi/2)+ \beta(\mu+h/2)\r]\;,\,\ov a(v)=\exp\l[\phi_\bb(v-\rmi/2) + \beta(\mu-h/2)\r]\nn\; .
\ee
The quantity $\delta$ in eqs. \refeq{beqv2}-\refeq{ceqv2} has been chosen
$\delta=1/2$ such that $a$, $\ov a$ are real valued functions. Define
$\lim_{\al\to\infty}(\al+1)J_\al =:J_0$. From the expression of $f_i$, eq. \refeq{ffromc} it then follows that the free energy
is that of an uncoupled impurity,
\be
\fl \lim_{\al\to\infty}f_i(T,h)=-T\ln\l[(a\ov a)^{-1}(u_0)+a^{-1}(u_0)+\ov
a^{-1}(u_0)\r]+2DJ_0-2\mu\nn\\
\lo=-T\ln\l[\te^{\beta 2DJ_0} + \te^{\beta(\mu+h/2)}+\te^{\beta(\mu-h/2)}\r]   \; .
\ee
The free energy reflects the expected result of a free impurity with three possible states,
namely empty and singly occupied with up or down spin. A more detailed
analysis of the case where $u_0$ is held fixed will be given in a forthcoming
publication \cite{akb}. We now demonstrate that
the two latter states dominate if $u_0$ is scaled appropriately with $\al, D,
\mu$, such that a crossover temperature emerges, below which a strong coupling fixed
point is reached. 

For low temperatures, the auxiliary function $\c$ exhibits a sharp
crossover from $\c\ll 1$ to $\c\gg1$ in regions around ''Fermi
points'' $\pm
\Lambda_\c$ defined by
\be
 -\phi_\c\l(\Lambda_\c\r)\approx 2\beta\mu\;,\qquad \Lambda_\c\approx\pm \al\sqrt{\frac D\mu-\frac14}\label{lamcdef}\;.
\ee
Set $h\ll \mu$. The influence of $h$ on the Fermi points is neglected, since
it enters quadratically. 
The more common parameterization is 
\be
\frac{v}{\al}=\frac12\tan \frac{k}{2}\label{vkrel}\; ,
\ee
where $k$ is the wave-vector used in the Fourier representation of the
Hamiltonian in the preceding section (not the Fourier variable conjugate
to $v$). At $v=\Lambda_\c$, eq. \refeq{lamcdef}
is equivalent to 
\be
2D(\cos k_F+1)= \mu \label{kfdef}\; ,
\ee
which defines $k_F$ at constant $\mu$ at $D\gg T$, such that $\mu=\e_F$ ($\mu$
in turn is related to the
particle number $N$ by $k_F=\pi N/(2L)$. Then eq. \refeq{kfdef} yields
a relation between $\mu$ and $N$, independent of $T$ - this demonstrates that the formally
grand-canonical description is effectively canonical, because $D\gg T$). 
For the analysis of the NLIE, consider 
\be
k_{\c,\b}* \ln \C(v)=k_{\c,\b}*\ln \c(v)+\int_{|w|<\L_\c}k_{\c,\b}(v-w)\ln
\frac{\C(w)}{\c(w)}\d w\label{carean}\; .
\ee
To evaluate the second integrand, note that asymptotically, 
\be
\fl \frac\C\c=\frac{1}{a\ov a}(1+a)(1+\ov a)=\Or\l(\frac{1}{\b\bb}\r)\label{bbcc}\\
\fl \ln \frac{\C(v)}{\c(v)}=\l\{
\begin{array}{ll} \ln \l[\l(1+a(v)\r) \l(1+\ov
a(v)\r)\r]+\l[-\phi_\c(v)-2\beta\mu\r],& |v|<\Lambda_\c \label{casdev}\\
0, &|v|>\Lambda_\c 
\end{array}
\r. \; .
\ee
Consider the case $|v|<\Lambda_\c$ in eq. \refeq{casdev}. The first term is exponentially small, the second term
dominates. It is inserted into the second integrand on the rhs of eq. \refeq{carean} which shall be considered as a
next-leading correction compared to the driving terms in
eqs. \refeq{beqv2}-\refeq{ceqv2}. In view of eq. \refeq{ffromc}, the most interesting range of the spectral parameter is $v\sim v_0$. Set 
\be
|u_0|=|v_0/2|> \Lambda_\c\sim \alpha\label{v0range}\,,
\ee
as indicated in \refeq{jalrange}. Then one proceeds with eq. \refeq{carean}  
\be
k_{\c,\b}* \ln \C(v)\stackrel{|v|>\Lambda_\c}{\approx} k_{\c,\b}*\ln \c(v)+k_{\c,\b}(v)\int_{|w|<\L_\c}\ln
\frac{\C(w)}{\c(w)}\d w+\Or(1/v^4)\nn\; .
\ee
As an estimate for the second term on the rhs, one uses the leading term of eq. \refeq{casdev}:
\be
\fl\ln \frac{\C(v)}{\c(v)}=
\l[-\phi_\c(v)-2\beta\mu\r]+\Or\l(\exp[-\beta D]\r)\nn\\
\fl\int_{|w|<\L_\c}\ln
\frac{\C(w)}{\c(w)}\d w\approx 4\beta \al D\l[2\arctan
\frac{2\L_\c}{\al}-\frac{\mu\L_\c}{\al D}+\Or(1/\al)\r]=:\beta \kappa>0 \label{defkappa}\; .
\ee
The sub-leading order $\Or(1/\al)$ is neglected in the following, which is justified
rigorously below. The above defined quantity $\kappa$ is a monotonously decreasing function of $\mu/D$. Choose $\L_\c>0$ such that $\kappa>0$. Summarizing,
\be
k_{\b,\c}*\ln \C=k_{\b,\c}*\ln \c+\beta \kappa k_{\b,\c}\label{ceqc}\; ,
\ee
so that $\ln \C$ can be eliminated in eqs. \refeq{beqv2}-\refeq{ceqv2}. The Fourier
transforms of $\phi_{\b}$, $\phi_{\bb}$, $\phi_\c$, eqs. \refeq{phib},
\refeq{phic} are:
\be
\hat \phi_\b(k)&=&-\beta D(\al+1)\l\{\begin{array}{ll}
                 \te^{-\al/2 k}\, , &k\geq0\\
                 \te^{(\al/2+1)k}\,,&k<0
                 \end{array}\r.\nn\\
\hat \phi_\bb(k)&=&-\beta D(\al+1)\l\{\begin{array}{ll}
                 \te^{-(\al/2+1) k}\, , &k\geq0\\
                 \te^{\al/2k}\,,&k<0
                 \end{array}\r.\nn\\
\hat \phi_\c(k)&=&\hat \phi_\b(k)+\hat \phi_\bb(k)=-\beta D(\al+1)\l\{\begin{array}{ll}
                 \te^{-\al/2 k}\l(1+\te^{-k}\r)\, , &k\geq0\\
                 \te^{\al/2k}\l(1+\te^{k}\r)\,,&k<0
                 \end{array}\r.\label{cfoutemp}
\ee
Inserting eq. \refeq{ceqc} gives, using eqs. \refeq{beqv2}-\refeq{ceqv2}:
\numparts
\label{auxfou}
\be
\hat \c(k)&=&\l\{\begin{array}{ll} 
              \frac{\hat\phi_\c}{1+\te^{-k}}-\frac{1}{1+\te^k} \beta \kappa -\frac{\hat\BB+\te^{-k}\hat\B}{1+\te^{-k}}& k\geq0\\
            \frac{\hat\phi_\c}{1+\te^{k}}-\frac{1}{1+\te^{-k}}\beta \kappa -\frac{\hat\B+\te^k\hat\BB}{1+\te^k}&k<0
            \end{array}\r.\label{auxfou1}\\
\hat\b(k)&=&-\frac{1}{1+\te^{-k}}\beta \kappa+\frac{1}{1+\te^{|k|}}(\hat\B-\hat\BB)\label{auxfou2}\\
\hat\bb(k)&=&-\frac{1}{1+\te^{k}}\beta \kappa+\frac{1}{1+\te^{|k|}}(\hat\BB-\hat\B)\label{auxfou3}
\ee
\endnumparts
Note that in the limit $\L_{\c}\to0$, i.e. $\mu\to 4D$, the resulting equations are trivially solved; $\ln \b=\beta h=-\ln \bb$, and
$\B=1+\exp(\beta h)$, $\BB=1+\exp(-\beta h)$, 
\be 
f_i={\rm const.} + \ln \l(\te^{\beta h/2}+\te^{-\beta h/2}\r)\label{htfe}\; ,
\ee
which is the free energy of a free, uncoupled spin. This situation corresponds to a completely filled band.  

The
NLIE eq. \refeq{auxfou1}-\refeq{auxfou3} are transformed back to direct space,
\numparts
\label{subeq7}
\be
\fl \ln \b(v)=-2\pi\beta\kappa \Phi(v+\rmi\delta) +\beta
h/2+[k*\ln\B](v)-[k*\ln \BB](v+2\rmi\delta)\label{heis1}\\
\fl\ln \bb(v)=2\pi\beta\kappa\Phi(v-\rmi\delta)-\beta
h/2+[k*\ln\BB](v)-[k*\ln \B](v-2\rmi\delta)\label{heis2} \\
\fl\ln \c(v)=  -\beta D(\al+1)\frac{\al}{v^2+\al^2/4}-k(v)\beta \kappa+\beta
\mu\nn\\
\lo +
[\Phi*\ln\B](v-\rmi\delta)-[\Phi*\ln\BB](v+\rmi\delta)\label{heis3}\; .
\ee
\endnumparts
The driving term and integration kernel read:
\be
\Phi(v)&=&\frac{\rmi}{2\sinh\pi v} \nn\\
k(v)&=& \frac{1}{2\pi} \i
\frac{\te^{-|k|/2}}{2\cosh k/2} \te^{\rmi kv} \d k\nn\; .
\ee
Choose $\delta=1/2$ and scale $v$ by
$1/\pi$. Since $\Phi$ decays exponentially, it is possible to absorb $\kappa$
and $v_0$ in a new additive constant. 
Substitute 
\be
v&=&x-\ln(2\pi\kappa)\label{shiftv}
\ee
and remember that $\kappa$ scales with $\al D$ \refeq{defkappa} and
therefore may be arbitrarily large. All parameters can be combined in the free
energy to a new constant $T_K$, 
\be
 -\ln(2\pi\kappa)-\pi v_0/2=: -\ln T_K\;,\qquad T_K= 2\pi\kappa\te^{\pi v_0/2}\label{tkdef}\; .
\ee
The range of $|v_0|$ has been identified in eq. \refeq{v0range}, we take
$v_0=-|v_0|$. The shift \refeq{shiftv} scales the driving term $\Phi(v/\pi-\rmi/2)$:
\be
-2\beta\pi\kappa\Phi(v/\pi -\rmi/2)&=& -\frac{\beta \pi\kappa }{\cosh v}\label{drivnonsc}\\
&=&
-\frac{2\beta \pi\kappa}{\te^{x-\ln(2\pi\kappa)}+\te^{-x+\ln(2\pi\kappa)}}\nn\\
&\stackrel{\kappa\to\infty}{=}&-\beta \te^x \nn\; .
\ee
In the second line, eq. \refeq{shiftv} has been employed. At this point it is clear that sub-leading orders in eq. \refeq{defkappa} can
safely be neglected. Furthermore, a factor $\beta$ can be absorbed by shifting $x\to x-\ln
\beta$:
\numparts
\label{subeq8}
\be
\ln \b(x)&=& -\te^{ x}  +\beta h/2+[k*\ln
\B](x)-[k*\ln\BB](x+\rmi\pi-\rmi \e)\label{bcon}\\
\ln \bb(x)&=& -\te^{ x} -\beta h/2+[k*\ln
\BB](x)-[k*\ln\B](x-\rmi\pi+\rmi\e)\label{bbcon}\\
-\beta f_i&=&\frac{1}{2\pi} \i \frac{[\ln \B\BB](x)}{\cosh\l(x+ \ln\frac{T}{T_K} \r)}\d
x \label{evcon}\; .
\ee
\endnumparts
As far as the host is concerned, eqations \refeq{auxlimff} are inserted into
equation \refeq{evhub}:
\be
-\lim_{\al\to\infty}\beta f_h&=& \eta(0)+\frac{1}{2\pi}\int_{\Lambda_\c/\al}^{\Lambda_\c/\al} 
\frac{1}{v^2+1/4}\ln\l[(1+a(v))(1+\ov a(v))\r]\d v\nn\; .
\ee
Observe the relation between the elementary excitation energy $\e(v)$ and
the momentum $k(v)$ as functions of the spectral parameter $v$ (cf. eq. \refeq{vkrel}),
\be
\e(v)=\frac{\d}{\d v} k(v)\label{ep}\;. 
\ee
Then one obtains the energy-momentum relation $\e(k)$: 
\be
\e(v)&=& \frac{1}{v^2+1/4}\; \rightarrow k(v)=2\arctan
2v\nn\\
\e(k)&=& 4\cos^2\frac{k}{2}=2\cos k+2\label{disp}
\ee
The function $k(v)$
in the first line is given by eq. \refeq{ep}. From eqs. \refeq{defkappa}, \refeq{tkdef}, $D$ gets arbitrarily large, leading to a linear dispersion in the host. Then the free energy {\em density} $f_h$ of the host is given by: 
\be
\fl f_h=- \lim_{\beta \gg 1}\frac{T}{2\pi} \int_{-k_F D}^{k_FD} \ln\l[1+\te^{-\beta
  (\widetilde v_F \l|q\r| -h/2-\widetilde \mu)}\r]\l[1+\te^{-\beta
  (\widetilde v_F\l|q\r|+h/2-\widetilde \mu)}\r]\d q\label{freeenho}\hspace{1cm}\\
\lo{\widetilde \mu:=}\mu-2D\l(\l(1+ \frac{3}{\al}\r)( \cos k_F+1)-\frac{2}{\al}(\cos
k_F+1)^2\r)\hspace{1cm}\nn\\
\lo{\widetilde v_F:=}2\l(1+\frac3\al-\frac4\al(\cos k_F+1)\r)\sin k_F\hspace{1cm}\nn\; ,
\ee
where $q=k\cdot D$. 
Note
that interactions in the host of order $\Or(1/\al)$ can be absorbed into a redefinition of $v_F=2
\sin k_F$, resulting in effectively free fermions. The leading orders of the specific heat and magnetic susceptibility are
\be
C_h(T)&=&T\,\frac{\pi^2}{3}\,\frac{2}{\pi \widetilde
  v_F}=:T\,\frac{\pi^2}{3}\,\widetilde \rho_h\label{hf1}\\
\chi_h(T)&=&\frac{1}{4} \,\frac{2}{\pi \widetilde v_F}=:\frac14 \widetilde \rho_h\label{hf2}\; ,
\ee 
where $\widetilde
\rho_h$ is the density of states in the host.\footnote[1]{The host in our model, eq. \refeq{defhhlindis}, contains two channels. Therefore $\tilde \rho_h$ is enhanced by a factor 2. The impurity couples only to one of the two channels.}

\section{Calculation of high- and low-temperature scales}   
We demonstrate the advantage of our novel approach by a direct calculation of high- and low-temperature scales and comparison with Wilson's results \cite{wil75}. 

Wilson found for $h,T\ll T_K$, the ratio of the specific heat $C_i$ to the magnetic susceptibility
$\chi_i$ is universal:
\be
\chi_i(T)&=&\frac{1}{2\pi T_K}\;,\;\;C_i(T)=\frac{\pi
  T}{3 T_K}\nn\\
R_w&:=&\frac{\chi_i(T)}{C_i(T)}\,\frac{C_h(T)}{\chi_h(T)}=2\label{defltwr}\;
.
\ee
We confirm these results in our approach by using dilogarithmic identities \cite{kl91b} to extract the lowest
order of the free energy for low fields and temperatures,
\be
\lim_{T,h\ll T_K}f(T,h)&=&-\frac{T^2}{2T_K}\,\frac{\pi }{3} -
\frac{1}{4\pi}\,\frac{h^2}{T_K}\label{lowft}\; .
\ee
From eq. \refeq{lowft},
\be
\lim_{T,h\ll T_K}C(T)&=&\frac{T}{T_K}\,\frac{\pi }{3}\qquad \lim_{T,h\ll
  T_K}\chi(h)=\frac{1}{T_K}\,\frac{1}{2\pi}\label{chidl} \; .
\ee
This Fermi liquid behavior is to be compared with the host, eqs. \refeq{hf1},
\refeq{hf2}. Define the coefficient of the linear $T$-dependence of $C_{h}$ by
$\delta_{h}$. Then $R_w$ is calculated as
\be
R_w&:=&\lim_{T\to 0}\frac{\chi}{\chi_h} \,
\frac{\delta_h}{\delta}=2\label{defwrlt}\; .
\ee
For high $T$ and $h=0$, the impurity susceptibility asymptotically reaches the
Curie-Weiss limit:
\be
\chi(T)= \frac{1}{4T}\l[1-\frac{1}{x}-\frac{\ln x}{
  2x^2} +\Or\l(x^{-3}\r)\r]_{x=\ln T/\widetilde T_K}\nn\; .
\ee
$\widetilde T_K$ is defined such that contributions $\Or\l(\ln^{-2}T/\widetilde T_K\r)$
do not occur,
\be
\widetilde T_K&=&2\pi \xi T_K\;,\qquad \xi=0.1032\pm 0.0005\label{defhtwr}\; ,
\ee
where the numerical value of $\xi$ was calculated by Wilson \cite{wil75}. The
calculation of the numbers $R_w$, $\xi$ are benchmarks for any
non-perturbative solution of the Kondo model covering the entire
temperature axis. In \cite{and80}, a perturbative expansion of the free
energy by Andrei gave an
analytical expression of $\xi$, $\xi=0.102676\ldots$. This
number has not been obtained in the
framework of the exact TBA solution yet. However, by analyzing the NLIE
eqs. \refeq{bcon}-\refeq{evcon} we are now able to give an accurate numerical
value of $\xi$, which is summarized in the rest of this section. 

In the high-temperature regime $T\gg T_K$, the asymptotic behavior of the auxiliary
functions for $x\sim -\ln T/T_K\ll 0$ gives the main contribution according to
eq. \refeq{evcon}. The convolutions in eq. \refeq{bcon} are evaluated
asymptotically at $h=0$: 
\be
{\rm Re }\;\ln \B(x\ll 0)&=& \ln 2 +\frac{\pi^2}{4\,x^3}\nn\\
{\rm Re }\;\partial_{\beta h} \ln \B(x\ll 0)&=&\frac{1}{2} +
\frac{1}{4\,x}-\frac{\ln
  |x|}{8x^2}+\frac{\phi}{x^2}\label{bmas}\\
{\rm Re }\;\partial^2_{\beta h} \ln \B(x\ll 0)&=&\frac{1}{4}\, \l(1+\frac{1}{x}-\frac{\ln |x|}{2
x^2}+\frac{4\phi+\frac{1}{4}}{x^2}\r)\nn\;.
\ee
The coefficient $\phi$ of the $x^{-2}$-decay in eq. \refeq{bmas} is determined
numerically; we find $\phi=0.04707\pm2\cdot10^{-7}$. From this one gets $\xi=\exp\l[-4\phi-1/4\r]/2\pi=0.102678\pm2\cdot
10^{-6}$, agreeing nicely both with Wilson's and Andrei's results. Details of the calculations will be published in a
forthcoming paper \cite{akb}.

\section{Conclusion and outlook}
An Anderson-like impurity Hamiltonian on a one-dimensional lattice has been obtained
as the logarithmic derivative of a gl(2$|$1) symmetric transfer matrix. The free energy for the lattice model was calculated
exactly from a closed set
of finitely many NLIE. As a special case, this impurity model allows for the Kondo limit of a localized magnetic impurity in an
interacting host of electrons including the free fermion case. Mathematically, the Kondo limit constitutes the reduction of gl(2$|$1)
symmetry to su(2) symmetry on the impurity site.

We expect that the lattice regularization presented in this work can be
generalized to the Kondo model with anisotropic
exchange by using the quantum super-algebra
U$_q$gl(2$|$1). Furthermore, it is possible \cite{frah99} to find representations of
gl(2$|$1) such that one of its subalgebras is the $(2S+1)$-dimensional
irrep of su(2). This allows for a lattice path integral approach to 
multichannel spin-$S$ Anderson-like models. In a forthcoming paper \cite{akb}, we will
give the corresponding closed set of NLIE in the Kondo limit. These can be obtained
by an argument of symmetry, and allow for the calculation of Wilson ratios in
the general multichannel spin-$S$ case. After having found a lattice path
integral approach in the simplest case (spin-$1/2$, single channel), the next
question addresses two-point correlation functions like $\langle
\sigma^\mu_i\sigma_{r}^{\mu'}\rangle,\;r=1,\ldots,L$. Since the spectrum and the
corresponding eigenstates of $T_{h,i}^{(Q)}$ are known, those quantities can be
calculated in principle, by generalizing methods developed in \cite{kit99} to
graded models. In view of progress in describing the
screening cloud around the impurity at $T\ll T_K$ by methods of conformal
field theory \cite{bar96, bar98} complementary understanding from the exact
solution is highly desirable. Finally, the Anderson impurity model on the
lattice, including charge fluctuations, can be treated exactly by
solving eqs.~\refeq{beqv2}-\refeq{ceqv2}, a question currently
under investigation. 

We acknowledge very useful discussions with F. G\"ohmann. 

\appendix
\section{Alternative fermionization and gl(2$|$1)-symmetry}
\label{appa}
\subsection{Alternative fermionization}
The matrices $X$ and $Y$ defined in \refeq{tildeX}, \refeq{Y} provide one possible fermionic representation of the $\l[m_j\r]_a^b$, $\l[e_0\r]_a^b$. This representation is not unique: In this appendix, we use slightly modified matrices $\widetilde X$, $\widetilde Y$. However, the fermionic representation of the Hamiltonian is essentially the same. Consider
\be
\widetilde X_j&=&\l(\begin{array}{cccc}
          (1-n_{j\downarrow})(1-n_{j\uparrow}) &
          (1-n_{j\downarrow})c_{j\uparrow} & c_{j\,\downarrow}(1-n_{j\uparrow})
          & c _{j\down}c_{j\up}\\
          (1-n_{j\down})c^\dagger_{j\up} & (1-n_{j\down})n_{j\up} &
          -c _{j\down}c^\dagger_{j\up} & -c _{j\down}n_{j\up}\\
          c^\dagger_{j\down}(1-n_{j\up})&c^\dagger_{j\down}c _{j\up}&n_{j\down}(1-n_{j\up})&n_{j\down}c_{j\up}\\
          -c^\dagger_{j\down}c^\dagger_{j\up}
          &-c^\dagger_{j\down}n_{j\up}&n_{j\down}c^\dagger_{j\up}&n_{j\down}n_{j\up}
       \end{array}\r)\label{XX}\; .
\ee
The first row and column are deleted to obtain
\be
\widetilde Y&=& \l(\begin{array}{cc|c}
          (1-n_{d,\down})n_{d,\up} &
          -d _{\down}d^\dagger_{\up} & -d_{\down} n_{d,\up}\\
         d^\dagger_{\down}d_{\up}&n_{d,\down}(1-n_{d,\up})&n_{d,\down}d_{\up}\\
          \hline -d^\dagger_{\down}n_{d,\up}&n_{d,\down}d^\dagger_{\up}&n_{d,\down}n_{d,\up}\\
       \end{array}\r)\nn\; .
\ee
Again one may identify $\l[m_j\r]^a_b=\l[\widetilde X_j\r]^a_b$,
$\l[e_0\r]_a^b=\l[\widetilde Y\r]_a^b$, since eqs. \refeq{comprop}, \refeq{projprop}
are fulfilled. The matrices $X$, $\widetilde X$ are related by a particle-hole transformation, so that fermionization of $H_h$ yields the same expression as
in eq. \refeq{hamhost}. The fermionic representation of the impurity operator is
denoted by $\widetilde H_i$:
\be
\fl \widetilde H_i=-\mu n_{d}+\frac
h2\l(n_{d,\up}-n_{d,\down}\r)-2 DJ_\al n_{d,\up}n_{d\down}\l(\al+\widetilde F_{1,L}\r)\nn\\
\lo  - DJ_\al\,\sqrt\al\,\sum_\tau
\l[d^\dagger_\tau(1-n_{d,\ov\tau})(c_{L,\tau}+c_{1,\tau})-d_\tau(1-n_{d,\ov\tau})(c^\dagger_{L,\tau}+c^\dagger_{1,\tau})\r]\nn\\
\lo  - DJ_\al \sum_\tau
n_{d,\tau}\l(2 n_{1,\ov\tau}+2n_{L,\ov\tau}-c^\dagger_{L,\ov\tau}c_{1,\ov\tau}-c^\dagger_{1,\ov\tau}c_{L,\ov\tau} \r)\nn\\
\lo  + DJ_\al\sum_\tau d^\dagger_\tau d_{\ov\tau}
\l(2c^\dagger_{L,\ov\tau}c_{L,\tau}+2c^\dagger_{1,\ov\tau}c_{1,\tau} -
c^\dagger_{L,\ov\tau}c_{1,\tau}-c^\dagger_{1,\ov\tau}c_{L,\tau}\r)+\Or\l(D J_\al/\al^{1/2}\r)\nn\;,
\ee
where
\be
\fl 2 F_{1,L}=2(n_1+n_L)-c^\dagger_{1,\up} c_{L,\up}-c^\dagger_{1,\down} c_{L,\down}-c^\dagger_{L,\up} c_{1,\up}-c^\dagger_{L,\down} c_{1,\down}\nn\; .
\ee
The canonical transformation generated by $A$, eq. \refeq{defA}, is also
applicable to $\widetilde H_i$, and the doubly occupied state is energetically suppressed by scaling $u_0$. In the Kondo limit, $H_i$ and $\widetilde H_i$
are seen to be identical.
\subsection{gl(2$|$1)-symmetry}
From the YBE \refeq{ybedef}, the direct product of two monodromy matrices is
intertwined by an $R$-matrix,
\be
\fl(-1)^{p[\beta'](p[\al]+p[\al'])}\l[T^{(4,4)}(u)\otimes
T^{(3,4)}(v)\r]^{\beta,\al}_{\beta',\al'}\,\l[R^{(3,4)}(v-u)\r]^{\al',\beta'}_{\al'',\beta''}\nn\\
\lo=\l[R^{(3,4)}(v-u)\r]^{\al,\beta}_{\al',\beta'}\,\l[T^{(3,4)}(v)\otimes T^{(4,4)}(u)\r]^{\al',\beta'}_{\al'',\beta''}(-1)^{p[\beta'](p[\al']+p[\al''])}\label{tmatcom0}\;.
\ee
The invariance of $\tau(u)$ with respect to gl$(2|1)$ is shown by expanding
eq. \refeq{tmatcom0} in the limit $v\to\infty$, keeping only terms $\Or(1)$,
$\Or(1/v)$.  
\be
\fl R^{(3,4)}(v)\sim1 +\frac1v\l(\frac\al2+1+(-1)^b e_a^bE_b^a\r)+\Or\l(\frac{1}{v^2}\r)\label{r34sim}\\
\fl T^{(3,4)}(v)=:
R^{(3,4)}_{a,L}(v)R^{(3,4)}_{a,L-1}(v)\ldots R^{(3,3)}_{a,0}(v+\rmi u_0)\nn\\
\lo\sim
1+\frac1v\l\{\sum_{j=1}^L\l[\frac\al2+1+(-1)^b\l[e_a\r]^b_a\l[E_j\r]^a_b\r]
+(-1)^b\l[e_a\r]^b_a\l[e_0\r]^a_b\r\}+\Or\l(\frac{1}{v^2}\r)\nn\\
\lo{=:} 1+\frac{W}{v} +\Or\l(\frac{1}{v^2}\r)\label{t34sim}\; .
\ee
$T^{(4,4)}(u)\equiv T(u)$ is defined in eq. \refeq{mondrom1}. 

Eqs. \refeq{r34sim}, \refeq{t34sim} are inserted into eq. \refeq{tmatcom0}
with $d''=3; d=d'=4$, while keeping the full $T(u)$. The constant
terms on both sides are identical. In order $\Or(1/v)$, one gets
\be
\fl(-1)^{p[\beta''](p[\al]+p[\al''])}[T(u)]^\beta_{\beta''}W^\al_{\al''}
+[T(u)]^\beta_{\beta''}\l[(-1)^{p[a]p[b]}e^b_aE^a_b\r]^{\al,\beta'}_{\al'',\beta''}\nn\\
\lo=(-1)^{p[\beta](p[\al]+p[\al''])}W^\al_{\al''}[T(u)]^\beta_{\beta''}
+\l[(-1)^{p[a]p[b]}e^b_aE^a_b\r]^{\al,\beta}_{\al'',\beta'}[T(u)]^{\beta'}_{\beta''}\nn\;
.
\ee
Set $\beta=\beta''$, multiply with $(-1)^\beta$ and sum over $\beta$. The
second terms on each side are identical. The first terms give the
commutator of the transfer matrix $\tau(u)$ with $W$: 
\be
\tau(u):=\sum_\beta(-1)^\beta [T(u)]^\beta_\beta\;,\,\l[\tau(u), W^\al_{\al''}\r]=0\label{comtauw}\; .
\ee
Dropping constants in $W$, eq. \refeq{comtauw} states that:
\be
\l[\tau(u), \sum_{j=1}^L\l[E_j\r]^a_b+\l[e_0\r]^a_b\r]=0\nn\; .
\ee
Thus $\tau$ commutes with all global gl(2$|$1) symmetry operators. In a very
similar way, one starts with eq. \refeq{ybebardef} to show
\be
\l[\ov\tau(u), \sum_{j=1}^L\l[E_j\r]^a_b+\l[e_0\r]^a_b\r]=0\nn\; ,
\ee
where $\ov \tau(u)$ is defined in eq. \refeq{bartaudef}. Consequently, the
Hamiltonian, defined by eq. \refeq{defham}, is gl(2$|$1)-symmetric.

\section{Derivation of NLIE}
\label{appb}
The unknown functions in eqs. \refeq{ev}-\refeq{defc} are $q_+,\;q_-,\;,q_-^{(h)}\;,q_+^{(h)}$, where the indices pertain to two different sets of particle and hole solutions. Incidentally for the largest eigenvalue, the index denotes the part of the
complex plane where these functions have zeroes: If a $q-$ function carries an
index $+$ ($-$), it has zeroes in the upper (lower) half plane. From numerical studies for finite $N$, we know that in the largest
eigenvalue case, the ``particle
solutions'' obey $\im[v_j]>0\;,\;
\im[\tilde v_k]<0
\;\forall j,k$. Analogously, the ``hole solutions'' are distributed in the complex plane as $\im[v_j^{(h)}]<1$,
$\im[\tilde v_k^{(h)}]>1\; , \forall j,k$. The particle and hole solutions are
distributed discretely and 
accumulate at certain points,
prohibiting a formulation in terms of densities. 

In the following, the largest eigenvalue case is studied. Consider the
logarithmic derivative of these auxiliary functions, so that constant terms vanish. Since we know the analyticity properties of all functions
in the complex $v$-plane, we can calculate their Fourier-transforms,
\be
\hat f=\i \l[\ln f(v)\r]'\,\te^{-\rmi kv} \frac{\d v}{2\pi} \label{foundef}\; .
\ee
The integration contour is taken along the real axis. This is allowed as long
as $|\al/2|>|u_N|$, which certainly is the case for $N$ and $\al$ sufficiently
large. $\hat f$ vanishes for $k<0$ ($k>0$) for $f(v)$ analytic
in $\mathds{C}^+$ ($\mathds{C}^-$). Thus it is convenient to calculate the
Fourier transforms separately for $k<0$, $k>0$. For the moment, concentrate on
$k<0$.
\be
-\hat \b(k) &=& -\te^{(\al/2+1)k}\hat\phi_-(k)+\te^k\hat q_+(k)\label{b}\\
-\hat \bb(k)&=& -\te^{k\al/2}\hat\phi_-(k)-\te^k\hat q_++\hat q_+^{(h)}\label{bb}\\
-\hat \c(k)&=& \te^{k\al/2}(\hat\phi_+(k)-\hat\phi_-(k))-\te^{(\al/2
  +1)k}\phi_-(k)+\hat\L_{i}(k)\label{c}\\
\hat\B(k)&=&-\te^k\hat q_+(k)+\hat\L_{i}(k)\label{B}\\
\hat\BB(k)&=&\hat q_+(k)-\hat q_+^{(h)}(k)+\hat\L_{i}(k)\label{BB}\\
\hat \C(k)&=& -\te^{k\al/2}\hat\phi_+(k)+\hat q_+^{(h)}(k)-\hat \L_{i}(k)\label{C}\; .
\ee
The essential observation is that in eqs. \refeq{b}-\refeq{C}, there appear
the three unknowns, namely $\hat q_+$, $\hat q_+^{(h)}$ and $\hat \L_{i}$, and
the three auxiliary functions $\hat \b$, $\hat \bb$ and $\hat \c$. Add eqs. \refeq{BB}, \refeq{C} and combine this sum with eq. \refeq{b}:
\be
\hat \b(k)=\te^{(\al/2+1)k}(\hat\phi_-(k)-\phi_+(k))
-\te^k(\hat\BB(k)+\hat\C(k))\label{bres}\; .
\ee
Combine eqs. \refeq{B} with \refeq{b} and these two with eq. \refeq{bres}. An
expression for $\hat \L_{i}$ results,
\be
\hat\L_{i}(k)&=& \hat \B(k)+\te^k(\hat\BB(k)+\hat
\C(k))+\te^{(\al/2+1)k}\hat\phi_+(k)\nn\; ,
\ee
which is inserted into eq. \refeq{c}:
\be
\fl\hat \c(k)= -\te^{k \al/2}(\hat
\phi_+(k)-\hat\phi_-(k))+\te^{(\al/2+1)k}(\hat\phi_-(k)-\hat\phi_+(k))-\hat
\B(k)-\te^k(\hat\BB(k)+\hat \C(k))\nn\;.
\ee
Finally, eqs. \refeq{B} and \refeq{C} give $\hat q_+^{(h)}(k)$, which is
inserted into eq. \refeq{bb},
\be
\hat\bb(k)&=&\te^{k \al/2} (\hat\phi_-(k)-\hat\phi_+(k))-(\hat \C(k)+\hat
\B(k))\; .
\ee
The case $k>0$ is obtained by exchanging $\hat \b$, $\hat \bb$, switching
$k\to-k$ in the exponential terms and replacing $\phi_-\leftrightarrow
\phi_+$. The result is summarized:
\be
\fl\hat \b(k)=\l\{\begin{array}{ll}
                \te^{-k \al/2}(\hat\phi_+(k)-\hat\phi_-(k))-(\hat\BB(k)+\hat
                \C(k))& k>0\\
                -(\hat\BB(k)+\hat
                \C(k))& k=0\\
                \te^{(\al/2+1)k}(\hat\phi_-(k)-\hat\phi_+(k))-\te^k(\hat\BB(k)+\hat \C(k)) & k<0
                \end{array} \r.\nn\\
\fl\hat \bb(k)=\l\{\begin{array}{ll}
                \te^{-(\al/2+1) k}(\hat\phi_+(k)-\hat\phi_-(k))-\te^{-k}(\hat\B(k)+\hat
                \C(k))& k>0\\
                -(\hat\B(k)+\hat
                \C(k))& k=0\\
                \te^{k \al/2}(\hat\phi_-(k)-\hat\phi_+(k))-(\hat\B(k)+\hat \C(k)) & k<0
                \end{array} \r.\nn\\
\fl\hat \c(k)= \l\{\begin{array}{ll}
                \l(\te^{-(\al/2+1)k} +\te^{-k\al/2
                }\r)(\hat\phi_+(k)-\hat\phi_-(k))-\hat\BB(k)-\te^{-k}(\hat\B(k)+\hat
                \C(k))& k>0\\
                -\hat\BB(k)-(\hat\B(k)+\hat \C(k))&k=0\\
                \l(\te^{(\al/2+1)k} +\te^{k \al/2}\r)(\hat\phi_-(k)-\hat\phi_+(k))-\hat\B(k)-\te^{k}(\hat{\overline{{\mathfrak{B}}}}(k)+\hat \C(k))& k<0
                \end{array}\r.\nn
\ee
Application of the inverse Fourier transform and integration leads to a system of non-linear
integral equations.
\be
\fl\ln \b(v)&=&\phi_\b^{(N)}(v+\rmi\delta) -[k_\b*\ln\BB](v+2\rmi\delta)-[k_\b*\ln\C](v+\rmi\delta)+\beta(\mu+h/2)\label{beq}\\
\fl\ln \bb(v)&=&\phi_\bb^{(N)}(v-\rmi\delta) -[k_\bb*\ln\B](v-2\rmi\delta)-[k_\bb*\ln\C](v-\rmi\delta)+\beta(\mu-h/2)\label{bbeq}\\
\fl\ln\c(v)&=&\phi_\c^{(N)}(v)
-[k_\b*\ln\BB](v+\rmi\delta)-[k_\bb*\ln\B](v-\rmi\delta)-[k_\c*\ln\C](v)+2\beta\mu\label{ceq}\;,
\ee
where the convolution 
\be
[f*g](x):=\i f(x-y)g(y)\d y\label{defconv}
\ee
is done with local kernels:
\be
\fl k_\b(v)=\frac{1}{2\pi v(v-\rmi)}\; ,\qquad k_\bb(v)=k_\b(v)^*\;,\qquad k_\c(v)=
k_\b(v)+k_\bb(v)=\frac{2}{2\pi(v^2+1)}\nn\; .
\ee
In order to achieve convergence, the equation for $\ln \b$ ($\ln \bb$),
eq. \refeq{beq} (eq. \refeq{bbeq}), is
taken for $v+\rmi\delta$, ($v-\rmi\delta$). 

The constant terms are integration constants derived from the asymptotic
behavior of the auxiliary functions for large $|v|$. 
\be
\lim_{|v|\to\infty}\b=\frac{a}{1+\ov a}\,,\;\lim_{|v|\to\infty}\bb=\frac{\ov
  a}{1+a}\,,\;\lim_{|v|\to\infty}\c=\frac{a\ov a}{1+a+\ov a}\label{asp}\\
a=\te^{\beta(\mu+h/2)}\; , \; \ov a=\te^{\beta(\mu-h/2)}\label{avdef}\;.
\ee
The inhomogeneities are
\be
\phi_\b^{(N)}(v)&=& \ln \frac{\phi_+\l(v+\rmi\frac\al2\r)\,\phi_-\l(v-\rmi\frac\al2-\rmi\r)}{\phi_-\l(v+\rmi\frac\al2\r)\,\phi_+\l(v-\rmi\frac\al2-\rmi\r)}\nn\\
\phi_\bb^{(N)}&=& \l[\phi_\b^{(N)}\r]^* \nn\\
\phi_\c^{(N)}&=&\phi_\b^{(N)}+\phi_\bb^{(N)}\nn\; .
\ee
The thermodynamic limit $N\to\infty$ can be carried out leading to eqs. \refeq{beqv2}-\refeq{ceqv2}.


\section*{References}
\begin{thebibliography}{99}
\bibitem{and61} Anderson P W 1961 {\it Phys. Rev.} {\bf 124} 41
\bibitem{sch66} Schrieffer J R and Wolff P A 1966 {\it Phys. Rev.} {\bf 149} 491 
\bibitem{kon64} Kondo J 1964 {\it Prog. Theor. Phys.} {\bf 32} 37
\bibitem{wil75} Wilson K G 1975 \RMP {\bf 47} 773
\bibitem{andr80} Andrei N 1980 \PRL {\bf 45} 379
\bibitem{and80} Andrei N and Lowenstein J H 1981 \PRL {\bf46} 356
\bibitem{wie81} Wiegmann, P B 1981 {\it J. Phys. C} {\bf 14} 1463
\bibitem{tsv83} Tsvelick A M and Wiegmann P B 1983 {\it Adv. Phys.} {\bf 32} 453
\bibitem{hal81} Haldane F D M 1981 \JPA {\bf 14} 2585
\bibitem{akb} Bortz M and Kl\"umper A 2004 submitted to EPJB 
\bibitem{bra95} Bracken A, Gould M D, Links J R and Zhang, Y Z 1995 \PRL {\bf 74} 2768
\bibitem{goe98} G\"ohmann F and Murakami S 1998 \JPA {\bf 31} 7729
\bibitem{goe00} G\"ohmann F and Korepin V E 2000 \JPA {\bf 33} 1199
\bibitem{goeh03} G\"ohmann F private communication
\bibitem{goe02} G\"ohmann F 2002 {\it Nucl. Phys. B} {\bf 620} 501
\bibitem{kldiss} Kl\"umper A PhDthesis (Universit\"at zu K\"oln) 1988
\bibitem{bar95} Bariev R Z, Kl\"umper A and Zittartz J 1995 {\it Europhys. Lett.}
  {\bf 32} 84
\bibitem{zvy97} Zvyagin A A and Schlottmann P 1997 J. Phys. Cond. Mat. {\bf 9} 3543; (E) 1997 {\bf 9} 6479
\bibitem{suz87} Suzuki M and Inoue M 1987 {\it Prog. Theor. Phys.} {\bf 78} 787
\bibitem{jue97} J\"uttner G, Kl\"umper A and Suzuki J 1997 \JPA {\bf 30} 1881
\bibitem{sak01} Sakai K and Kl\"umper A 2001 \JPA {\bf 34} 8015
\bibitem{kl91b} Kl\"umper A, Batchelor M T and Pearce P A (1991) \JPA {\bf 91}
  3111
\bibitem{bar96} Barzykin V and Affleck I 1996 \PRL {\bf 76} 4959 
\bibitem{bar98} Barzykin V and Affleck I 1998 {\it Phys. Rev. B} {\bf 57} 432
\bibitem{frah99} H. Frahm; Nucl. Phys. B 559 (1999) 613
\bibitem{kit99} Kitanine N, Maillet J M and Terras V 1999 {\it Nucl. Phys. B}
  {\bf 554} 647
\end{thebibliography}
\end{document}